\documentclass[aps,prd,twocolumn,superscriptaddress,floatfix,showpacs]{revtex4}
\usepackage{graphics}
\usepackage{graphicx}
\usepackage{subfig}
\usepackage{epsfig}
\usepackage{amsmath}
\usepackage{amsfonts}
\usepackage{bm}
\usepackage{url}
\usepackage{float}
\usepackage{multirow}
\usepackage{xcolor}
\usepackage{xspace}
\usepackage{listings} 
\usepackage[normalem]{ulem}

\def\beq{\begin{equation}}
\def\eeq{\end{equation}}
\def\bea{\begin{eqnarray}}
\def\eea{\end{eqnarray}}

\def\dd{\text{d}}

\def\msun{\,{\rm M_\odot}}
\def\gsim{ \lower .75ex \hbox{$\sim$} \llap{\raise .27ex \hbox{$>$}} }
\def\lsim{ \lower .75ex\hbox{$\sim$} \llap{\raise .27ex \hbox{$<$}} }

\def\th{\theta}
\def\a{\alpha}
\def\b{\beta}
\def\m{\mu}
\def\e{\epsilon}

\def\s{\sigma}
\def\k{\kappa}
\def\l{\lambda}
\def\L{\Lambda}

\def\lm{\lambda^{\m}}

\def\lk{\lambda^{\k}}

\def\tjn{\theta_{JN}}
\def\ctjn{\cos\tjn}

\def\mc{\mathcal{M}}
\def\lnmc{\ln\mathcal{M}}
\def\lnmu{\ln\mu}

\def\lnL{\ln\mathcal{L}}
\def\dk{\partial_\k}

\def\dd{\text{d}}

\def\num{num}

\def\tf2{\texttt{TaylorF2} }
\def\imrpd{\texttt{IMRPhenomD} }
\def\imrpdnrt{\texttt{IMRPhenomD-NRTidal} }

\def\beq{\begin{equation}}
\def\eeq{\end{equation}}
\def\bea{\begin{eqnarray}}
\def\eea{\end{eqnarray}}

\definecolor{dark_green}{HTML}{427A10}

\newcommand{\dhmc}{{\sc DeepHMC}\xspace}

\def\exp{exp}

\begin{document}
\input epsf.tex

\title{\dhmc : a deep-neural-network acclerated Hamiltonian Monte Carlo algorithm for binary neutron star parameter estimation}
\author{Jules \surname{Perret}}
\email[]{perret@apc.in2p3.fr}
\author{ Marc \surname{Ar\`ene}}
\email[]{marc.arene@apc.in2p3.fr}
\author{Edward K. \surname{Porter}}
\email[]{porter@apc.in2p3.fr}
\vspace{1cm}
\affiliation{Universit\'{e} Paris Cit\'{e}, CNRS, Astroparticule et Cosmologie, 75013 Paris, France}
\vspace{1cm}
\begin{abstract}

We present a deep neural network  (DNN) accelerated Hamiltonian Monte Carlo (HMC) algorithm called 
\dhmc  for the inference of binary neutron star systems.  The HMC is a non-random walk sampler that uses background gradient information to accelerate the convergence of the sampler.  While faster converging than a random-walk sampler, in theory by a factor of the dimensionality of the problem,  a known computational bottleneck for HMC algorithms is the calculation of gradients of the log-likelihood.  We demonstrate that Hamiltonian trajectories based on a DNN gradients are 30 times faster than those based on the relative binning gradients, and 7000 times faster than trajectories based on a naive likelihood gradient calculation.  Using the publicly available 128 second LVK data set for the binary neutron star mergers GW170817 and GW190425,  we show that  not only does  \dhmc produce produces highly accurate and consistent results with the LVK public data, but acquires 5000 statistically independent samples (SIS) in the $12D$ parameter space in approximately two hours on a Macbook pro for GW170817, with a cost of $<1$ second/SIS, and 2.5 days for GW190425, with a cost of $\sim25$ seconds/SIS.  
\end{abstract}

\maketitle

\section{Introduction}
Since the first discovery of gravitational waves (GWs) from the coalescence of a binary black hole (BBH) in 2015~\cite{Abbott:2016blz}, the LIGO-Virgo-Kagra collaboration (LVK) has announced the detection of 90 high significance events in total~\cite{LIGOScientific:2018mvr, LIGOScientific:2020ibl, LIGOScientific:2021djp}.  Amongst these events, which are mainly composed of the coalescence of BBHs, there is a growing population of sources which are most likely to be from the coalescence of either binary neutron star (BNS)~\cite{TheLIGOScientific:2017qsa, LIGOScientific:2020aai} or neutron star - black hole (NSBH) systems~\cite{LIGOScientific:2021qlt}.

To extract the astrophysical parameters from GW events, the LVK uses Bayesian inference algorithms based on matched filtering. Under the assumption of Gaussian noise, the matched filter is the optimal linear filter for extracting signals buried in noise. This powerful technique uses theoretical GW templates to phase match potential signals in the data, allowing for their detection and identification.  Within the framework of Bayesian inference, the posterior density distribution for the source parameters is given by Bayes' theorem.  However,  due to the high dimensionality of the problem, a direct solution to Bayes' theorem is not available.  In general, stochastic samplers are used to approximate the posterior density.  At present, the LVK uses variants of Markov chain Monte Carlo (MCMC) ~\cite{hastings_1970,metropolis_1953} or nested sampling~\cite{skilling_2006} algorithms to extract the parameters of the system~\cite{Ashton:2018jfp, Speagle_2020} 
These algorithms belong to a family of stochastic, random walk samplers, which are known to have potentially slow convergence properties~\cite{2011hmcm.book..113N}. 

As the current 2G detectors improve their low frequency performance, and with future 3G detectors such as Einstein Telescope~\cite{Punturo_2010, abac2025scienceeinsteintelescope} and Cosmic Explorer~\cite{reitze2019cosmicexploreruscontribution,evans2023cosmicexplorersubmissionnsf}, we expect to see an increase in the signal durations, the event signal-to-noise
ratio (SNR) and the number of detectable events.  These will have a direct impact on the convergence of current samplers, resulting in a large increase in the computational resources needed for future inference~\cite{hu2025costsbayesianparameterestimation}.  In recent years, 
accelerating Bayesian inference for GW events has been an active field of research.  At each step of the sampler an evaluation of the likelihood function, which normally requires the generation of an expensive gravitational waveform
(or template), is required.  To improve the runtime for parameter estimation samplers, different groups have focused on different aspects, such as accelerating the likelihood computation~\cite{cornish2013fastfishermatriceslazy, 2018arXiv180608792Z, Morisaki_2021, Canizares_2015, Smith_2016}, marginalizing over extrinsic
parameters~\cite{lange2018rapidaccurateparameterinference, morisaki2023rapidlocalizationinferencecompact, Roulet_2024}, applying machine learning techniques~\cite{Williams_2021, Dax_2025}, using metric or gradient-based samplers~\cite{Porter:2013wwa, Bouffanais:2018hoz, nitz2024robustrapidsimplegravitationalwave}, or combinations of the above~\cite{wong2023fastgravitationalwaveparameter}


 In this work, we present a gradient-based Hamiltonian Monte Carlo (HMC) algorithm~\cite{duane_1987} for BNS inference, where the gradients are provided by a deep neural network(DNN) model.  The HMC was initially developed for lattice field simulations and sought to suppress the random walk behaviour normally associated with stochastic samplers.  Instead of proposing random jumps in parameter space, the HMC evolves trajectories based on Hamiltonian dynamics in phase-space that are driven by the gradient of the target distribution.  The HMC is, in theory, superior to standard MCMC algorithms as it uses information from the background geometry of the parameter space and, as a consequence, is able to explore distant points faster than other MCMC methods. If the algorithm is well tuned, the acceptance rate of the HMC algorithm is usually very high and the autocorrelation between adjacent points of the chain is very low, giving many more SISs for the same number of chain points. Empirically, studies have shown that the HMC is approximately $D$ times more efficient than standard MCMC samplers, where $D$ is the dimensionality of the parameter space \cite{Hajian:2006mt,Porter:2013wwa}. 
 

While known to be more efficient, the HMC is not as commonly used as other samplers.  There are two main reasons for this: firstly,  the algorithm has a number of 
free parameters that are difficult to tune.  Secondly,  the HMC treats the inverted target distribution as a ``gravitational" potential well.  While simulating the dynamics of the trajectory in phase-space, each
successive step requires the calculation of $D$-dimensional gradients of the target density.  If there
is no closed-form solution for the gradients, as is the case in GW astronomy, they have to be calculated using expensive numerical differencing techniques which involve generating multiple waveform
templates,  resulting in trajectory generation times that are too slow for practical application.  This leads to a major computational bottleneck and can result in a total runtime that makes the HMC impractical for use.

A potential solution to the gradient bottleneck was proposed for high signal to noise ratio (SNR) GW sources observable with the LISA detector~\cite{Porter:2013wwa}.  In that work, the HMC algorithm was broken into three separate phases.
Phase I was a  purely information gathering phase, where a relatively small number of trajectories, i.e. $\sim{\mathcal O} (10^3)$, were generated using numerical derivatives.  As the algorithm traversed the target density, for each accepted trajectory, the
visited points and their $D$-dimensional gradients were recorded.  This information was then used in Phase II to derive the coefficients of an analytical cubic-fit approximation to the gradients of the log-likelihood.   Assuming
 that the information gathered in Phase I was sufficiently accurate, and that the cubic gradient approximation was a good representation of the numerical gradients, this allowed us to create a ``shadow"
 potential which had the same form as the target density.  In Phase III, the Hamiltonian at the beginning of each trajectory was calculated numerically using waveform templates.  The algorithm then moved from the true potential
 to the shadow potential in order to simulate the Hamiltonian trajectories using analytic rather than the expensive numerical gradients.  At the end of the trajectory, we moved back to the true potential to again numerically calculate the end-point Hamiltonian.   Given that the analytic gradient trajectories were on the order of ${\mathcal O}(10^3)$ times faster to generate, this resulted in a viable alternative to more standard inference algorithms.
 
 While this method worked for high SNR LISA sources, where the target distribution is unimodal,  when applying the HMC to the inference of BNS sources in LVK data, we found that the cubic gradient 
 approximation failed when the posterior distributions were multimodal~\cite{Bouffanais:2018hoz}.  In order to circumvent this problem, we introduced a dual approach to calculating the gradients of the 
 log-likelihood.  For parameters with unimodal posterior distributions, we continued to use the cubic gradient approximation.  For parameters with multimodal distributions, we introduced the use of 
 ordered lookup tables (OLUTs).  This again involved recording information on visited points and the log-likelihood gradients, creating ordered tables for each of the parameters, and then using a local
 linear approximation to predict the gradients of a new point.  While not as fast as the cubic approximation, the combination of the cubic-OLUTs method still allowed us to greatly speed up the inference.
Using a TaylorF2 waveform model~\cite{Buonanno_2009}, and a 9$D$ parameter space that neglected spins and tidal deformation of the bodies, we demonstrated that this timescale could be reduced to 1-10 secs / SIS using the HMC.

While this feasibility study was encouraging,  there were a number of outstanding issues.  Firstly, the study was based on a non-spinning waveform with no tidal interactions.  Secondly, the 
algorithm was tested on a small number of BNS sources consisting of simulated signals in simulated noise.   Finally, while the cubic-OLUTs gradient approximation worked well, it required us to know
a priori which parameters were unimodal and which were multimodal, something that is usually not known until the inference has already begun.

In this work,  we present a deep neural network (DNN) based HMC algorithm called \dhmc.   This algorithm improves on the HMC algorithm presented in~\cite{Bouffanais:2018hoz} in a number of ways.  Most significantly,
we replace the cubic-OLUTs gradients with gradient approximations generated by a DNN model.    This way, we no longer need to pre-define the modality of each parameter distribution and thus achieve an accelerated, and more precise, gradient approximation as compared to the cubic-OLUTs method .  We also
extend the algorithm to use waveform models containing spins and tidal deformation. As the LVK now uses a Python based 
library called Bilby~\cite{Ashton:2018jfp}, we have ported \dhmc from  C to Python, and are now fully compatible with Bilby.  

The rest of the article is organized as follows: In Sec. II we present the generic HMC algorithm and discuss how we tune the free parameters.  We also discuss how we deal
with parameter boundaries and diagnostic tools used to analyse the performance of the HMC.  In Sec. III we detail the extension of the parameter space to include
spins and tides.  In Sec. IV we discuss the inclusion of astrophysical priors and how we deal with the problem of calculating gradients of the log-prior distributions.  In 
Sec. V we discuss the construction of the deep neural network used in this study and present the final structure of the \dhmc algorithm.  In Sec. VI we apply \dhmc
to the 128 second public data for GW170817 and GW190425\cite{gwosc170817, gwosc190425}.  The former was a bright BNS event with SNR = 32, observed in all three LIGO and Virgo detectors.  For this source, all of the posterior densities are
unimiodal in nature.  The latter was a low-SNR (SNR = 12) event, observed in LIGO Livingston only, but which uses Virgo data for parameter estimation.  This source has multiple modes in many parameters.  This event provides a strong test of the algorithms abilitiy
to cope with multimodal structures on the likelihood surface.  In this section, we will present a comparison of results against the public LVK results.  Sec. VII concludes
 the paper with a discussion of future plans.


\section{Bayesian Inference for gravitational wave parameter estimation}
\label{sec:bayesian_inference}
\subsection{Introduction}
A GW impinging on a detector induces a time-domain strain, $s(t)$, which is a linear combination of the noise in the detector, $n(t)$, and a GW signal, $h(t)$,  i.e.
\begin{equation}
s(t) = h(t) + n(t),
\label{presentation_problem_data_analysis}
\end{equation}
where we assume that the noise is stationary and Gaussian.

Bayesian inference for GW astronomy involves using a data set $s(t)$, a theoretical waveform model $\tilde{h}(\lk$) (based on general relativity) described by the set of parameters $\lk$,
 and a model for the one-sided power spectral density (PSD) $S_n(f)$.  With these, one can construct the posterior probability distribution $p(\lk |s)$ via Bayes' theorem
\begin{equation}
p(\lm | s) = \frac{p(s | \lk) \pi(\lk)}{p(s)},
\label{Bayes_theorem}
\end{equation}
where $p(s | \lk) = \mathcal{L}(\lk)$ is the likelihood function (defined below), $\pi(\lk)$ is the prior distribution reflecting the a-priori knowledge of our parameters and $p(s)$ is the marginalized likelihood or ``evidence" given by
\begin{equation}
p(s) = \int \mathcal{L}(\lambda^{\k}) \pi (\lambda^{\k})  \dd \lambda^{\k}.
\end{equation}
Under the assumption of Gaussian noise, the likelihood in Eq.~(\ref{Bayes_theorem}) is defined as~\cite{Finn:1992wt}
\begin{equation}
\mathcal{L}(\lambda^{\k}) \propto \exp \left[ - \frac{1}{2} \left< s - h(\lambda^{\k}) | s - h(\lambda^{\k})\right>  \right],
\label{eq:likelihood}
\end{equation}
where the angular brackets denote the noise-weighted inner product,
\begin{equation}
\left< h | g \right> = 2 \int_{0}^{\infty} \frac{\tilde{h}(f)\tilde{g}^{*}(f) + \tilde{h}^{*}(f)\tilde{g}(f)}{S_{n}(f)}  df.
\label{eq:Scalarproduct}
\end{equation}
Here $\tilde{h}(f)$ is the Fourier transform of the time-domain waveform and the asterisk represents a complex conjugate. In this study, we use the LIGO and Virgo PSDs associated with the public release of the GW170817 data files~\cite{gwosc170817, gwosc190425}.


\subsection{Hamiltonian Monte Carlo}
\label{sec_description_HMC}
Using the concept of a canonical distribution from statistical mechanics,  given an energy function $E(\vec{x})$, the canonical distribution over states
has a probability density function
\beq
P(\vec{x}) = \frac{1}{Z} \exp\left(\frac{-E(\vec{x})}{T}\right)=\frac{1}{Z} \exp\left(\frac{-\mathcal{H}(q^{\k},p^{\k})}{T}\right),
\eeq
where $Z$ is a normalising constant, $\mathcal{H}(q^{\k},p^{\k})$ is an invariant Hamiltonian function defined in terms of the position and momentum state space variables $(q^{\k},p^{\k})$ and $T$ is the temperature of the system (which we take as being $T=1$).  We define the Hamiltonian
\beq
H(q^{\k},p^{\k}) =U(q^{\k}) + K(p^{\k}),
\eeq
composed of a potential energy term,  $U(q^{\k})$, and a kinetic energy term, $K(p^{\k})$,    thus defining a joint density
\beq
P(q^{\k},p^{\k}) = \frac{1}{Z}\exp(-U(q^{\k})) \exp(-K(p^{\k})).
\eeq

In Bayesian inference, we are most interested in the posterior distribution for the model parameters.  With this in mind, we equate the state space position variables with the waveform model parameters,
i.e. $q^{\k}=\lambda^{\k}$, and define a potential energy term
\beq
U(q^{\k}) = - \ln \left[ \pi(q^{\k}) {\mathcal L}(q^{\k}) \right],
\eeq
where again $\pi(q^{\k})$ is the parameter prior distribution and $\mathcal{L}(q^{\k})$ is the likelihood function.

While the potential energy term is completely determined by the target distribution, the distribution for the kinetic energy term has no constraint, and is open to definition by the user.  In most applications of the
HMC algorithm, it is common to use a quadratic form for the kinetic energy term, i.e.
\begin{equation}
K(p^{\k}) = \frac{1}{2}M_{\k\s}^{-1}p^{\k}p^{\s},
\label{kinetic_energy}
\end{equation}
where $M_{\k\s}$ is a fictitious positive-definite mass matrix.  The advantage of this choice is that the kinetic energy term is now the negative log
probability density of a zero-mean Gaussian distribution with covariance matrix $M_{\k\s}$.  If we further specify that the mass matrix is diagonal, we can
define a vector of mass scalings $m_{\k} = M_{\k\k}$.  This allows us to specify that each of
the momenta are independent of each other, with each component having zero mean and variance $m_{\k}$.  We can now define the kinetic energy term as
\beq
K(p^{\k}) = \sum_{\k=1}^{D}\frac{(p^{\k})^2}{2m_{\k }}.
\eeq

With this definition, we can further write the joint density as
\beq
P(q^{\k},p^{\k}) = P(q^{\k})P(p^{\k}) = P(q^{\k})\mathcal{N}(0,m_{\k}).
 \label{Separable_proba}
\eeq
Not only is the above density separable, but by construction,  the momentum components are independent of the state space variables
$q^{\k}$ and each other. Thus by sampling $(q^{\k},p^{\k})$ from the above distribution, and by simply ignoring the momentum variables, the marginal distribution for $q^{\k}$ gives us a sample set which asymptotically comes from the target distribution~\cite{2011hmcm.book..113N}.

As the Hamiltonian $H(q^{\k},p^{\k}) $ defines the total energy of a system, the dynamical evolution of the system in fictitious time $t$ can then be inferred from Hamilton's equations,
\begin{eqnarray}
\frac{\dd q^{\k}}{\dd t} &=& \frac{\partial H}{\partial p^{\k}} = \frac{\partial K}{\partial p^{\k}}, \label{theoretical_Hamil_eq_1}  \\
\frac{\dd p^{\k}}{\dd t} &=& -\frac{\partial H}{\partial q^{\k}} = -\frac{\partial U}{\partial q^{\k}}.
\label{theoretical_Hamil_eq_2}
\end{eqnarray}

In the most basic implementation of the HMC, the positions and momenta are evaluated discretely along a trajectory of total length $L$, composed of $l$ steps with step-size $\epsilon$, i.e. $L=l\e$.   The problem with this formulation is that as $\epsilon$ is
a constant in all dimensions, the algorithm does not take the different dynamical scales in the parameter coordinates into account.  To compensate for this, we use a range of step sizes $\epsilon^\k$ (defined below),  each individually tailored to a specific parameter.  In solving any Hamiltonian system, we need to ensure that the Hamiltonian is conserved along the trajectory, that the trajectories are time reversible and that the phase-space volume is conserved.  This is achieved using a symplectic integrator (commonly referred to as a leapfrog integrator).  Assuming our mass matrix is represented by a
diagonal matrix with elements $m_\k$, we can write the scaled leapfrog equations as~\cite{2011hmcm.book..113N}
\begin{eqnarray}
\tilde{p}^{\k}\left(\tau + \epsilon^{\k}/2\right) &=& \tilde{p}^{\k}(\tau) - \frac{\epsilon^{\k} }{2} \frac{\partial U(q^{\k}) }{\partial q^{\k}} \Bigr|_{q^{\k}(\tau)}, \nonumber \\
q^{\k}(\tau + \epsilon^{\k} ) &=& q^{\k}(\tau) + \epsilon^{\k} \tilde{p}^{\k}\left(\tau + \frac{\epsilon^{\k}}{2}\right),  \label{eq:Scaled_leapfrog}\\
\tilde{p}^{\k}(\tau + \epsilon^{\k} ) &=& \tilde{p}^{\k}\left(\tau + \frac{\epsilon^{\k}}{2}\right) - \frac{\epsilon^{\k} }{2} \frac{\partial U(q^{\k}) }{\partial q^{\k}} \Bigr|_{q^{\k}(\tau + \epsilon^{\k})}\nonumber ,
\end{eqnarray}
where we define the scaled momenta $\tilde{p}^{\k} = s^{\k}p^{\k}$ and the scaled step sizes $\epsilon^{\k} = s^{\k} \epsilon$, with $s^{\k} = m^{-1/2}_{\k}$. Without loss of generality, the scaled momenta are also drawn from a normal distribution, $P(\tilde{p}^{\k}) \sim \mathcal{N}(0,1)$.  When using the scaled momenta, the kinetic energy term now
reduces to
\beq
K(p^{\k}) = \frac{1}{2}\sum_{\k=1}^{D}(\tilde{p}^{\k})^2.
\eeq

As we are solving
Hamilton's equations discretely, the Hamiltonian is not strictly conserved along the trajectory.  In general, the leapfrog method introduces errors in the Hamiltonian on the order of $\mathcal{O}(\epsilon^{3})$ at each step of the trajectory and of order $\mathcal{O}(\epsilon^{2})$ over the whole trajectory.   To conserve the Hamiltonian, the values of $\epsilon^{\k}$ should be chosen to be as small as possible to ensure conservation, but
not so small as to introduce a new computational bottleneck into the algorithm.
To make the HMC ``exact", a Metropolis-Hastings evaluation is introduced at the end of each trajectory.

The HMC algorithm then works as follows: beginning with a point in parameter space, $q^{\k}$, we
\begin{enumerate}
\item Draw the scaled momenta $\tilde{p}^{\k}$ from a Gaussian distribution $\mathcal{N}(0,1)$,
\item Now, starting at the phase-space point $(q^\k_i, \tilde{p}^\k_i)$ with Hamiltonian $H(q^\k_i, \tilde{p}^\k_i)$, we evolve the Hamiltonian trajectory for $l$ steps to a new phase-space coordinate $(q^\k_f, \tilde{p}^\k_f)$ with Hamiltonian $H(q^\k_f, \tilde{p}^\k_f)$,
\item Accept or reject the new phase-space point with probability $\alpha = min\left(1,\exp\left[ -H(q^\k_f, \tilde{p}^\k_f) + H(q^{\k}_i,\tilde{p}^{\k}_i)\right]\right)$.
\end{enumerate}
As we stated above, the development of a HMC algorithm requires a greater overhead cost as compared to other stochastic samplers.  This is due to the free parameters 
within the algorithm which are difficult to tune, and from the computational bottleneck produced by the need to calculate the $D$-dimensional gradients of the target density at each step of the trajectory, especially when no analytic closed
form solution exists.  Below, we treat both issues and offer potential solutions to create an efficient algorithm.


\subsection{Tuning the free parameters in the HMC}
As with all stochastic samplers, there are a number of free parameters within the algorithm that need to be tuned.  For the HMC used in this study, these are: the step size in the Hamiltonian
trajectories $\e$, the number of steps $l$ in each trajectory defining the trajectory length $L = l\e$, and the diagonal elements of the mass matrix $m_{\k}=M_{\k\k}$
which define the scales $s_{\k}=1/\sqrt{m_{\k}}$ for the momenta and step sizes. 

From empirical studies, we found that a step size of $\e = 5\times10^3$  works very well in Phase I, giving a good balance between exploration and acceptance rate.  However,    during the inference phase, to create a more efficient algorithm, we found that rather than using a constant step size, it was better to have step sizes drawn from
the distribution $\e = {\mathcal N}(5\times10^{-3}, 1.5\times10^{-3}) \in [10^{-3}, 10^{-2}]$~\cite{2011hmcm.book..113N,Bouffanais:2018hoz}.  The advantage of drawing $\e$ from a distribution rather than working with a constant value is twofold.  First, in general it is difficult to find a ``one size fits all" value.  A step size that works for BNS sources may not work for BBH sources.  Secondly, when drawing from a distribution, we are almost guaranteed that all trajectories based on small step sizes will be accepted, while all accepted trajectories using larger step sizes will provide a more global exploration of the target density.

\begin{figure}[t]
       \includegraphics[height=6cm, width=0.48\textwidth]{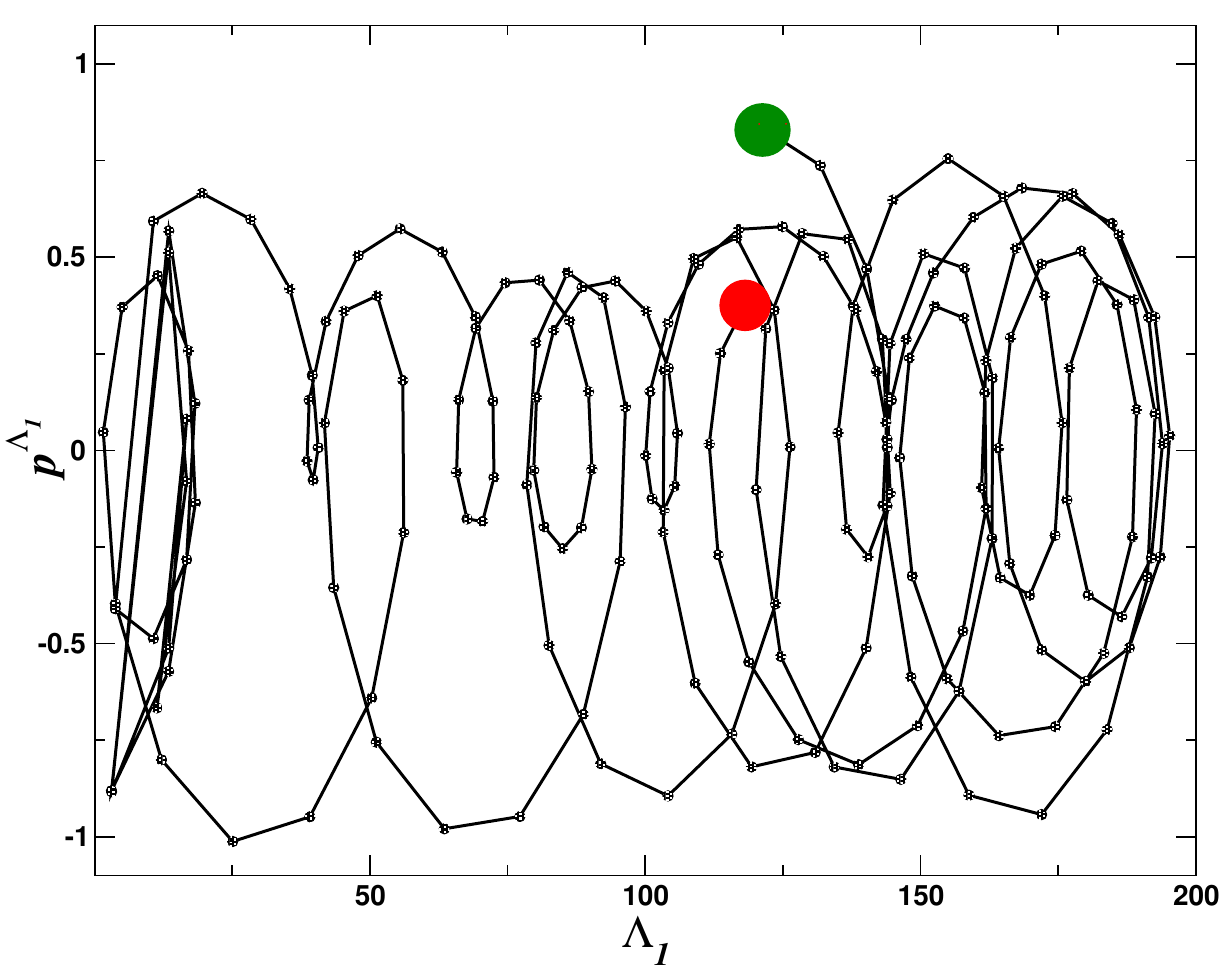}
        \caption{Phase space trajectory for the tidal parameter $\Lambda_1$ demonstrating how trajectories with a large number of steps can result in proposed points (red) very close to the initial point (green), leading to random walks in parameter space.}
        \label{fig:longl}
\end{figure}

Tuning the number of trajectory steps is a delicate process.  If $l$ is too small, the trajectory moves a short distance
in phase-space and the algorithm random walks through parameter space.  On the other hand, if $l$ is too big, we end up completing full orbits in phase-space with the 
likelihood at the end point of the trajectory not very far from the starting point, again causing the algorithm to random walk through parameter space.  In Fig.~\ref{fig:longl}
we provide an example of a phase space trajectory where the initial (blue) and proposed (red) points are very closely located in parameter space, even though the exploration
itself is large.  

The final free parameter to  tune is the mass matrix.  The HMC requires an initial ``guess" at how broad the target density is.  For this, we assume that the Fisher information matrix (FIM),
defined by
\begin{equation}
\Gamma_{\k\s} =  - \mathbb{E} \left[ \frac{\partial^{2} \ln \mathcal{L}}{\partial \lambda^{\k} \lambda^{\s}} \right] = \left< \frac{\partial \tilde{h}}{\partial \lambda^{\k}} \Big\rvert \frac{\partial \tilde{h}}{\partial \lambda^{\s}} \right>,
\label{FIM_statistics}
\end{equation}
is a good approximation to the local curvature of the log-likelihood surface, and is thus a good approximation to the width of the target density.  Given that the FIM is proportional to the metric tensor in the parameter space~\cite{Owen:1995tm}, it also provides 
information on the background geometry that assists the efficiency of the Hamiltonian trajectories.  We should mention here that while the FIM suffers from a number of issues, e.g. it
can often be singular, it has no knowledge of other modes etc., it is a viable option here as a first approximation to initially help the HMC move through parameter space.  With these 
assumptions, we define our scales as
\beq
s^{\k} \equiv m^{-1/2}_{\k} = C_{\k\k}^{1/2} = \sigma^{FIM}_{\k},
\eeq
where $C_{\k\s} = \Gamma^{-1}_{\k\s}$ is the variance-covariance matrix, and $\sigma^{FIM}_{\k} $ are the standard deviations predicted by the FIM.  As mentioned before, the FIM can
quite often be (almost) singular leading to standard deviation predictions that are larger than the natural and/or prior range for a particular parameter.  When this happens, we set the
scale in that parameter to have a width of 1/12 of the prior (coming from the assumption that if a parameter has a bimodal structure within the prior range, the distribution has essentially $12\sigma$ range).  We should highlight here that the use of the FIM matrix is only to get the HMC working in the initial information gathering phase.  At the end of this phase, we re-calculate the scales based on the covariance matrix of the accumulated data.  

Before progressing, we will take the opportunity to define the sampler-space coordinates to be used for the HMC.  For a coalescing binary, the gravitational waveforms are characterized by the parameter set $~{\l^{\k} = \{m_1, m_2, \vec{S}_1, \vec{S}_2, \Lambda_1, \Lambda_2, D_L, \a, \delta, \tjn, t_c, \varphi_c\\, \psi \}}$, where $(m_1, m_2)$ are the component masses, $(\vec{S}_1, \vec{S}_2)$ are the component spins, $(\Lambda_1, \Lambda_2)$ are the component tidal parameters, $D_L$ is the luminosity distance to the source, $(\a,\delta)$ are the right ascension and declination of the source, $\tjn$ is an inclination angle that we will define later in the article, $t_c$ is the GPS time of coalescence for an event, $\varphi_c$ denotes the phase of the waveform at coalescence and $\psi$ is the polarization angle.

Geometrically, if we were to directly equate the above parameter set $\l^{\k}$ with the sampler-space coordinates $q^\k$ it would create a highly degenerate and problematic coordinate system. To avoid issues like this, we demonstrated in previous studies that better parameterizations exist that provide more stable, non-degenerate coordinate systems~\cite{Cornish:2006ry, Cornish:2006ms}.  After testing multiple coordinate systems,  we define the following sampler-space coordinate system for \dhmc: $q^\k = \{\lnmc, \lnmu, \chi_1, \chi_2, \Lambda_1, \Lambda_2, \ln D_L, \alpha, \delta, \ctjn,\allowbreak \ln \delta t_c, \varphi_c, \psi  \}$.  Here ${\mathcal M} = n\eta^{3/5}$ and $\mu=m\eta$ are
the chirp mass and reduced mass respectively, where $m = m_1+m_2$ and $\eta=m_1 m_2/m^2$ define the total mass and symmetric mass ratio,  and $(\chi_1, \chi_2)$ are a set of dimensionless spins that we will define in Sec. IV.  Instead of using the time of coalescence, $t_c$, we found that algorithm is more efficient if
we use the duration of the signal, $\delta t_c$, instead.  We should highlight that using the logarithm of certain parameters allows us to reduce the condition number of the FIM and hence reduce numerical instabilities when calculating its inverse.

\subsection{Efficiently calculating the potential energy gradients in Phase I.}
While our ultimate goal is to use a DNN to model the gradients of the target density in Phase III, in Phase I we need to use a full numerical approximation to the gradients.  As shown above, the potential energy in the Hamiltonian is defined as
\beq
U(q^{\k}) = - \ln \left[ \pi(q^{\k}) {\mathcal L}(q^{\k}) \right],
\eeq
where $ \pi(q^{\k})$ is the prior distribution and ${\mathcal L}(q^{\k})$ is the likelihood as defined by Eqn~(\ref{eq:likelihood}).  If we expand this equation we find a data only term $\left< s | s \right>$, which we can
neglect for our purposes, allowing us to define the log-likelihood ratio
\beq
\ln\Lambda(q^{\k}) = \left< s | h \right> - \frac{1}{2}\left< h | h \right>.
\label{eq:likelihood_ratio}
\eeq
We now define the gradients of the potential energy by
\beq
\frac{\partial U(q^{\k})}{\partial q^{\k}} = -\frac{\partial \ln\pi(q^{\k})}{\partial q^{\k}} - \frac{\partial \ln{\Lambda}(q^{\k})}{\partial q^{\k}}.
\eeq
In previous works, we calculated the gradients of the log-likelihood ratio using numerical differencing of waveform templates, i.e.
\beq
\frac{\partial \ln\Lambda(q^{\k})}{\partial q^{\k}} = \left< s \left| \frac{\partial h}{\partial q^{\k}} \right>\right. - \left< h \left| \frac{\partial h}{\partial q^{\k}} \right>\right.
\eeq
However, as the waveforms have amplitudes on the order of $h\sim10^{-21}$, the numerical differencing can lead to errors in the log-likelihood derivative.  To circumvent this problem,
we now calculate the derivatives by numerically differencing the log-likelihood ratios directly, i.e.
\beq
\frac{\partial \ln\Lambda(q^{\k})}{\partial q^{\k}} = \frac{\ln\Lambda(q^{\k} + \Delta q^{\k}) - \ln\Lambda(q^{\k} - \Delta q^{\k})}{2 \Delta q^{\k}}.
\eeq
As well as being more numerically stable, by using the Bilby relative binning likelihood approximation~\cite{2018arXiv180608792Z}, we can accelerate the derivative calculation by a factor of $10^2$ over the direct waveform calculation.


\subsection{Dealing with prior and astrophysical boundaries in phase-space}
As with all Bayesian samplers, the algorithm is unaware of an physical boundaries that may exist in the problem.  In our case, we have two 
different boundary types that can be encountered.  The first are boundaries imposed by the astrophysical priors we use, while the second is due
to the special case where $m_1=m_2$.

To deal with the prior boundaries, we assume that a Hamiltonian trajectory impinges on a boundary from an initial phase-space state $(q_0, p_0)$
to a state $(q_n, p_{n+1/2})$, just inside the prior boundary, $q^{B}$.  Evolving the leapfrog equations another iteration moves the 
position of the trajectory to a new proposed value $q_{n+1}^{p} $ according to
\beq
q_{n+1}^{p} = q_n + \epsilon p_{n+1/2}.
\eeq
Without a boundary treatment, if the new point $q_{n+1}^{p} $ is outside the prior range, it will overshoot the boundary by an 
amount $\Delta q = \pm(q_{n+1}^{p}  - q^B$), depending on whether we are at the upper/lower prior boundary .  To prevent this from happening, we assume that the prior boundaries constitute an infinite energy
barrier that the Hamiltonian trajectories cannot penetrate.  We further impose a reflective boundary condition by assuming that the trajectory undergoes 
a perfectly elastic collision at the barrier.  As we now require the kinetic energy before and after the collision to be both constant in magnitude and opposite in sign,
this implies
\beq
\left|p_{(n+1)+1/2}\right| = -\left|p_{n+1/2}\right|.
\eeq
Correspondingly, the positional component is reflected back inside the prior boundary by the amount of overshoot, i.e. 
\beq
q_{n+1} = q^B \mp \Delta q = 2q^B - q_{n+1}^{p},
\eeq
again depending on whether we are at the upper/lower prior boundary.

We can use a similar argument to treat the equal mass boundary.  As each parameter in the leapfrog equations is updated individually, a mapping of the 
$m_1 = m_2$ line defines a minimum value of $\mc$, when $\m$ is held constant, beyond which we are into the unastrophysical part of the parameter
space which results in complex individual masses.  This relation is given by 
\beq
\mc^{min}(\mu) \approx 1.74\, \mu .
\eeq
Holding $\mu$ constant, if a Hamiltonian trajectory tries to go below $\mc^{min}(\mu)$ to a proposed point $\mc^p_{n+1}$, the trajectory again undergoes a perfectly elastic collision with the positional component being reflected to a point
\beq
\mc_{n+1} = 2\mc^{min}(\mu) - \mc^p_{n+1}.
\eeq
Similarly for $\m$, we can find an inverse relation that give $\m^{max}(\mc)$.  Keeping $\mc$ constant, any trajectory that tries to go beyond this value is
also reflected to the position
\beq
\m_{n+1} = 2\m^{max}(\mc) - \m^p_{n+1}.
\eeq
 In both cases, we also reverse the sign of the momentum to conserve the kinetic energy under the reflection.


\subsection{Diagnostic Tools}
A difficulty with all Markov chain methods is estimating both the convergence and convergence efficiency of a particular algorithm.  In the past, a popular metric was
the acceptance rate.  We know, theoretically, that an optimally designed MCMC or HMC algorithm will have desired acceptance rates of $~23\%$ and $~65\%$ respectively~\cite{2011hmcm.book..113N}.
However, given that a high acceptance rate is not a guarantee of accelerated convergence, it has been proposed (see for example~\cite{dewitt-morette_monte_1997}) that the 
effective (statistically independent) sample size, $N_{\text{ESS}}$, is a better metric.   We define the effective sample size as
\begin{equation}
N_{\text{ESS}} = \frac{N}{\tau_{int}},
\end{equation}
where $N$ is the total number of samples and $\tau_{\text{int}}$ is the integrated autocorrelation time (IAT) defined by
\begin{equation}
\tau_{\text{int}}(N) = 1 + 2 \sum_{\tau=1}^{\infty} \rho (\tau)\approx 1 + 2 \sum_{\tau=1}^{N} \rho (\tau),
\label{eq:iat}
\end{equation}
assuming a sufficiently large $N$.  Here,  $\rho(\tau)$ is the autocorrelation function at lag $\tau$, given by 
\begin{equation}\label{eq:autocorr}
\rho(\tau)  = \dfrac{\displaystyle\sum_{i=1}^{N - \tau} \left( X_{i} - \overline{X} \right) \left( X_{i + \tau} - \overline{X} \right)  }{\displaystyle \sum_{i=1}^{N } \left( X_{i} - \overline{X} \right)^{2}},
\end{equation}
where $X_i$ denote the chain samples and $\overline{X}$ denoting the sample mean. 
   
An advantage of using the effective sample size is the fact that it is directly related to the statistical error is the estimation of the posterior distribution~\cite{dewitt-morette_monte_1997}.  
This error scales as  $(\tau_{\text{int}}/N)^{1/2} = (1/N_{\text{ESS}})^{1/2}$, meaning that to have a 1\% accuracy in our estimate of the credible intervals, we need to have $N\sim10^4 \tau_{int}$, or in other words, around $10^4$ SISs.  

A common problem in calculating the IAT is knowing when to stop the integration.   As can be seen from Eq.~\eqref{eq:iat}, the evaluation of $\tau_{\text{int}}$ assumes a very large number of samples.  However, in general, the autocorrelation function drops off in an exponential  manner  reaching zero at lags of $\tau \ll N$.  Furthermore, when it does eventually reach zero, $\rho (\tau)$ continues to oscillate around zero as $N$ grows.  The consequence of this is that the variance of Eq.~\eqref{eq:iat} does not go to zero as $N\rightarrow\infty$ due to the additional "noise" in the integrand.  To circumvent this, we adopt the solution proposed by Sokal~\cite{dewitt-morette_monte_1997, Madras:1988ei}, where we to stop the integration at a sample size $M \ll N$, assuring the integral is still
dominated by the autocorrelation and not by the numerical noise.  This solution works (for sample sizes of $\geq 10^3$)  by searching for the smallest value of $M$ for which the inequality
\beq
M \geq C \tau_{\text{int}}(M)
\label{eq:iatstop}
\eeq
holds. It is usual to take $C = 5$ for autocorrelations with an exponential drop-off (such as we have in this study).



\section{Extending the HMC to include spins and tidal deformation}
While the original HMC was shown to work in the case of non-spinning binaries, it is of fundamental importance to include spin and tidal effects in the waveform for BNS systems.  In this study we
use the \imrpdnrt waveform~\cite{Dietrich:2017aum, Dietrich:2019kaq}, which uses the aligned spin \imrpd waveform~\cite{PhysRevD.93.044006, Khan:2015jqa} as its basis and includes estimation of tidal effects which have been calibrated with
numerical relativity simulations.

For coalescing binaries, spin effects appear to leading order at the 1.5 PN order through spin-orbit coupling, and then at the 2 PN order through spin-spin coupling~\cite{Apostolatos:1994mx}.  The spins are defined by the parameters
$\left(\vec{S}_1, \vec{S}_2\right)$ which are, in general, misaligned with the orbital angular momentum $\vec{L}$, leading to precessional effects.  If we assume, as is the case with the 
waveform model we are using, that $\left(\vec{S}_1, \vec{S}_2\right)$ are aligned with $\vec{L}$, then there is no precession and we can parameterize the spins in the dimensionless form
\beq
\chi_i = \frac{c\vec{S}_i.\vec{L}}{Gm_{i}^{2}},
\eeq
with $\chi_i\in[-1,1]$, and where $c$ is the speed of light and $G$ is Newton's constant.  Then we can capture the dominant spin effects in the aligned-spin waveform by defining the effective spin~\cite{Ajith:2009bn, Santamaria:2010yb, Damour:2001tu}
\beq
\chi_\text{eff} = \frac{m_1\chi_1 + m_2\chi_2}{m_1+m_2}.
\eeq
In general, if the component bodies are non-spinning, we can define the inclination angle as the angle between the line of sight unit vector $\hat{n}$ and $\hat{L}$, i.e. $\cos\iota = \hat{n}.\hat{L}$.  However, once we include spins, this angle is no longer constant and we can no longer use $\cos\iota$ as the inclination parameter.   In contrast, in the case of simple precession, the angle between $\hat{n}$ and the total angular momentum $\hat{J}$ is approximately constant during the binary evolution and 
allows us to use the parameter $\cos\theta_{JN} = \hat{n}.\hat{J}$ instead.

For stiff compact objects such as NSs, when the separation of the binary is small enough, each component in the binary system will begin to tidally deform the other~\cite{Flanagan:2007ix}.  The dominant tidal effects
appear at the 5 PN and 6 PN orders in the phase of the waveform~\cite{Wade:2014vqa}.  These effects are described by the dimensionless tidal parameters
\beq
\L_i = \frac{2}{3}k_2\left[\frac{c^2}{G}\frac{R_i}{m_i}\right]^5,
\eeq
where $k_2$ is the $l=2$ Love number and $R_i$ is the radius of the NS~\cite{Flanagan:2007ix}.  Within the LVK, it has been demonstrated that the random-walk inference algorithms work 
better using the mass-weighted combinations~\cite{Favata:2013rwa}
\bea
 \tilde{\L} &=& \frac{8}{13}\left[\left(1+7\eta-31\eta^2\right)\left(\Lambda_1+\Lambda_2\right)\right. \nonumber \\
 &+&\left. \sqrt{1-4\eta}\left(1+9\eta-11\eta^2\right)\left(\Lambda_1-\Lambda_2\right) \right],
\eea
and
\bea
\delta\tilde{\Lambda} &=& \frac12\left[\sqrt{1-4\eta} \left(1-\frac{1372}{1319}\eta+\frac{8944}{1319}\eta^2\right)(\Lambda_1+\Lambda_2) \right. \nonumber\\
          &+&\left. \left(1-\frac{15910}{1319}\eta+\frac{32850}{1319}\eta^2+\frac{3380}{1319}\eta^3\right)(\Lambda_1-\Lambda_2) \right].\nonumber\\
\eea
However, as it is believed that $\delta\tilde{\Lambda}$ is unmeasurable with current second generation detectors,  the inference is dominated by the single tidal paramater $\tilde{\L}$. With this in mind, and given that we are using the background geometry of the target distribution to move in parameter space, we found that it was sufficient to sample directly in $(\Lambda_1, \Lambda_2)$.  For this analysis, we will also assume that both NSs share the same equation of state.

By including spin and tidal parameters, we increase the dimensionality of the parameter space to $D=13$.  However, as it is unlikely that any higher harmonic
modes will be visible in BNS systems, we take advantage of the fact that we can marginalise over the phase at coalescence $\varphi_c$ and reduce the 
parameter space to $\bar{D}=12$.  A consequence of this choice is that we will  need to calculate the FIM on the ($D$-1) subspace in order to obtain the 
initial scales in the HMC.  To do this, we calculate the FIM on the $D$-dimensional space, and then project onto the orthogonal ($D$-1) subspace using~\cite{Owen:1995tm}
\beq
\gamma_{ij} = \Gamma_{ij} - \frac{\Gamma_{i\varphi_c} \Gamma_{\varphi_cj} }{\Gamma_{\varphi_c\varphi_c} },
\label{eqn:pmaxfim}
\eeq
where we use latin indices to clarify that we are working in a reduced dimensionality.  This projection both minimizes the proper distance between two points
on the subspace and maximizes the FIM over the projected parameter..

By definition, the initial Hamiltonian scales are derived from the FIM.  When including spin and tidal parameters, our initial acceptances rate in Phase I dropped to zero.  Upon investigation, we found that
once we included $\{\chi_i, \Lambda_i\}$, the FIM scales in $\{\lnmc,\lnmu,\ln t_c\}$ were orders of magnitude larger than in the non-spinning case and were causing the kinetic energy terms to blow up in the 
Hamiltonian.  Furthermore, we found that the scales in $\{\chi_i, \Lambda_i\}$ were much larger than their posterior distributions.  An easy workaround to this problem was the following: we calculate the
phase-maximized FIM in both the non-spinning with tides and spinning only scenarios, and simply take the smallest scales from each of the two FIMs.  
As usual, for any parameter with scales larger than the prior distribution, we simply set the scales to half the width of the prior.

While not mathematically correct, we remind the reader that the scales are only used to get the HMC moving efficiently through parameter space in Phase I, and that these scales are replaced
before Phase III begins using information derived directly from the Phase I chain points.


\section{Including astrophysical priors}


\subsection{Transforming the LVK priors}
In Bayesian inference, the choice of prior distributions can have a large impact of the speed of convergence to the target density.   
To be as consistent as possible with previous LVK analyses, we take the prior on distance to be quadratic in the range $D_L\in[10,100]$ Mpc.   As we are
using $\ln(D_L)$ as our sampling parameter, this transforms to $\pi(\ln D_L)=D_L^3$ prior.  Assuming isotropic distributions
of orientation and sky positions, we take $\{\cos\th_{JN}, \sin\delta\}\in[-1,1]$, and $\a\in[0, 2\pi]$.  The polarization angle is constrained to $\psi\in[0,\pi]$ and the tidal parameters are  taken to be uniform between $\Lambda_i\in[0,5000]$.  We take the prior on the signal duration to be uniform in the range $\pi(\ln t_d)= U[t_d - 0.5, t_d+0.5]$ secs.   

For the spin parameters, the LVK analysis involves a low-spin prior such that $\chi_i\in[-\chi_{max}, \chi_{max}]\equiv[-0.05, 0.05]$.  However, so that comparisons can be made with precessing approximants, 
it is necessary to consider uniform priors on the spin magnitude $\vec{S}_i$ which yields the so-called zprior~\cite{Lange:2018pyp}
\beq
\pi(\chi_i) = -\frac{1}{2\chi_{max}}\ln\left(\left| \frac{\chi_i}{\chi_{max}}\right|\right).
\eeq

For the mass prior, things are more complicated.  To sample the parameter space, the LVK analysis assumes a flat prior in the individual masses, i.e. $\pi(m_1, m_2) \propto 1$.  As \dhmc samples $(\lnmc, \lnmu)$
space, we need to transform this joint prior.  Beginning with the expression
\bea
\pi(\ln\mc, \ln\m) &=& J\, \pi(m_1, m_2) \\ \nonumber
&=&  \left| \frac{\partial(m_1, m_2)}{\partial(\ln\mc, \ln\m)}\right|\pi(m_1, m_2) ,
\label{eq:massprior}
\eea
where $J$ is the Jacobian determinant,  it can be shown (see Appendix A) that the Jacobian has the form
\beq
J(\mc,\m) = \frac{5\mc^{5/2}}{2\sqrt{\m}\sqrt{ 1-4\left(\frac{\m}{\mc}\right)^{5/2}}}.
\label{eq:jacobian}
\eeq
Using the fact that $(\m/\mc)^{5/2} = \eta$, we can see that  the Jacobian determinant scales as $J\propto 1/(1-4\eta)^{1/2}$, and is thus singular for equal mass
binaries.  To circumvent this behaviour, we replaced the 
singular analytic function with an approximation that asymptotically approaches the equal mass line.  To do this, we use a binomial series of the form
\beq
\frac{1}{(1-\alpha)^{\b}} = \sum_{k=0}^{\infty}\binom{\b+k-1}{k}\a^{k},
\eeq

\begin{figure}[t]
       \includegraphics[height=6cm, width=0.45\textwidth]{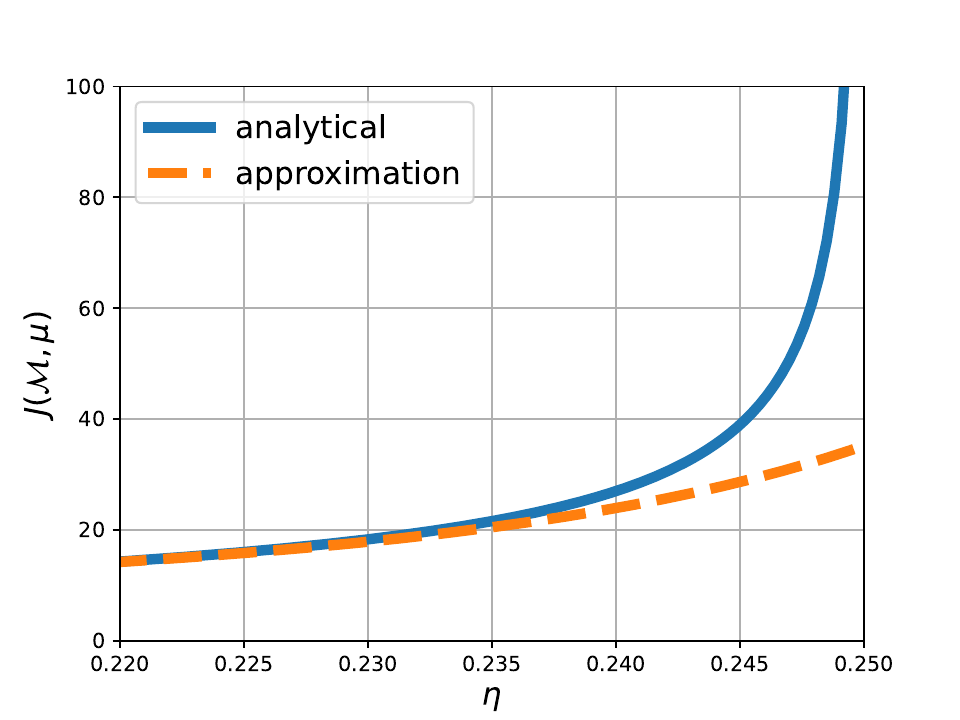}
        \caption{A comparison between the analytic (blue-solid line) and approximate (orange-dashed line) Jacobian determinant as a function of symmetric mass
        ratio, $\eta$.  As we approach the equal mass boundary, the analytic Jacobian diverges to infinity.  The approximate Jacobian avoids this behaviour, and smoothly transitions to the equal mass line.}
        \label{fig:Jacobian}
\end{figure}

which allows us to write the Jacobian as
\beq
J(\mc,\m) \approx \frac{5\mc^{5/2}}{2\sqrt{\m}}\sum_{k=0}^{n}\binom{1/2+k-1}{k}\left[4\left(\frac{\m}{\mc}\right)^{5/2}\right]^k.
\eeq
In Fig.\ref{fig:Jacobian} we plot both the analytic and approximate Jacobians.  We can see that while the analytic Jacobian (blue solid curve) diverges as
we approach $\eta=1/4$, the approximate Jacobian (orange dashed curve) smoothly approaches the equal mass limit.  Empirically we found that a limit of $n=30$
in the summation gave a good match to the analytic Jacobian at low values of $\eta$ and produced a smooth asymptotic curve at higher values.


\subsection{Gradients of the astrophysical priors}
Taking gradients of the potential energy requires us to also take gradients of the log-priors.  For the spin and mass priors, this is problematic as the gradients of
their log-priors display singular behaviour.  We discuss each in turn below and outline the solutions we propose to solve the problems.

\subsubsection{Gradients of the spin log-prior}

\begin{figure}[t]
       \includegraphics[height=7cm, width=0.45\textwidth]{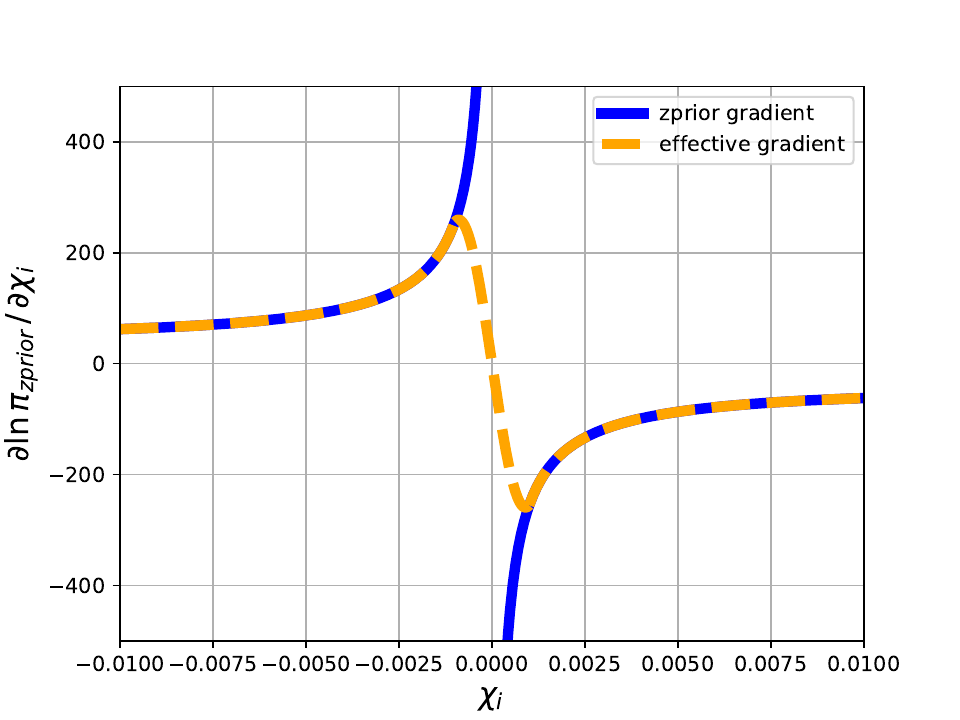}
        \caption{A comparison between the analytic (blue-solid line) and effective (orange-dashed line) gradients of the log-zprior as a function of the spin parameter $\chi_i$.  We see that the analytic gradient of the log-zprior diverges to $\pm\infty$
        as we approach $\chi_i=0$.   The effective log-zprior gradient smoothly connects the two sides of parameter space.}
        \label{fig:zpriorgrad}
\end{figure}

Taking the derivative of the spin log-prior, i.e.
\beq
\frac{\partial \ln \pi(\chi_i)}{\partial \chi_i} = \left[ \chi_i \ln \left(| \chi_i / \chi_{max}|\right)\right]^{-1},
\eeq
we arrive at an expression which causes multiple problems for the HMC.  We can see that the above expression contains singularities at $\chi_i = \pm\chi_{max}$ and $\chi_i=0$.
The boundary singularities are easy to solve by again applying a reflective boundary.  Empirically we found that setting this boundary at $\chi_B=\pm0.99\chi_{max}$ is sufficient
to prevent the HMC from failing, while still allowing exploration of the majority of the prior range.  The divergence at $\chi_i=0$ is more difficult to solve.  As we can see in Fig.~\ref{fig:zpriorgrad}, the gradient of the log-zprior diverges as we approach
$\chi_i=0$.  It is not possible to simply excise a small area around  $\chi_i=0$ as this would produce a potential energy barrier that the HMC would find impossible to cross.  To solve the problem, we implement a sinusoidal extrapolation function that smoothly connects the two sides of parameter space
through $\chi_i=0$ in the form
\begin{equation}
\frac{\partial \ln \pi(\chi_i)}{\partial \chi_i} = \left\{ \begin{array}{ll} \left[ \chi_i \ln \left(| \chi_i / \chi_{max}|\right)\right]^{-1}& \chi_{tr} < |\chi_i| < \chi_{max} \\ \\ A\sin\left(\frac{\pi\chi_i}{1.8\chi_{tr}}\right) & |\chi_i| < \chi_{tr} \end{array}\right.,
\end{equation}
where $\chi_{tr}=10^{-3}$ is the empirically determined spin value where we transition between the true and effective gradients, and the amplitude $A$ is given by
\beq
A = \left[ \chi_{tr} \ln\left(\frac{\chi_{tr}}{\chi_{max}}\right)\sin\left(\frac{\pi}{1.8}\right)\right]^{-1}.
\eeq
We can see in Fig.~\ref{fig:zpriorgrad} that this new effective log-zprior gradient now smoothly connects both sides of the spin parameter space allowing the HMC to smoothly transition across the $\chi_i=0$ singularity.


\subsubsection{Gradients of the mass log-prior}
Starting with the logarithm of Eqn.(\ref{eq:massprior}), we can define the gradient of the log-prior for the masses as
\beq
\frac{\partial\ln\pi(\ln\mc, \ln\m)}{\partial\ln q^{\mu}} = \frac{\partial \ln J}{\partial\ln q^{\mu}} .
\eeq
As we can see, the log-prior gradients involve gradients of the log-Jacobian determinant.  So, taking derivatives of the logarithm of Eqn.(\ref{eq:jacobian}), we obtain the expressions
\bea
\frac{\partial\ln\pi(\ln\mc, \ln\m)}{\partial\ln\mc} &=& \frac{5(\mc^{5/2}x - 2\m^{5/2})}{2\mc^{5/2}x}.\\
\frac{\partial\ln\pi(\ln\mc, \ln\m)}{\partial\ln\m}&=& -\frac{(\mc^{5/2}x - 10\m^{2})}{2\mc^{5/2}x}, 
\eea
where, for ease of expression, we define
\beq
x = 1 - 4(\m/\mc)^{5/2}.
\eeq
We can see once again, that each of the log-prior gradients are singular as we approach equal masses.  So again, we can avoid this singular behaviour by defining approximate gradients that asymptotically approach the equal mass boundary in a smooth fashion.   For the gradients of the log-prior, things are simpler as the binomial series for $1/(1-\a)$ reduces to a special case of the geometric series, i.e. 
\beq
\frac{1}{1-\a} = \sum_{k = 0}^{\infty} \a^k.
\eeq
Substituting this expression into the above equations allow us to write 
\bea
\frac{\partial\ln\pi(\ln\mc, \ln\m)}{\partial\ln\mc} &=&\frac{5(\mc^{5/2}x - 2\m^{5/2})}{2\mc^{5/2}}\sum_{k = 0}^{n} \left[4\left(\frac{\m}{\mc}\right)^{5/2}\right]^k,\nonumber\\ \\
\frac{\partial\ln\pi(\ln\mc, \ln\m)}{\partial\ln\m}&=&  -\frac{(\mc^{5/2}x - 10\m^{2})}{2\mc^{5/2}}\sum_{k = 0}^{n} \left[4\left(\frac{\m}{\mc}\right)^{5/2}\right]^k,\nonumber\\
\eea
where once again we found that a limit of $n=30$ in the summation is sufficient for our needs.


\section{A deep neural network approximation to the gradients of the log-likekihood}
Over the past decade, there has been an explosion in the development and application of machine learning (ML) algorithms.  One aspect of ML that is particularly suited to GW
inference is that of deep neural networks (DNNs).  DNNs attempt to recognize complex patterns in data using a generic scheme where several layers of simulated neurons are
stacked on top of each other.  Then instead of trying to use an a priori model to fit the data, as we were doing with the cubic-OLUTs approximation, the DNN uses stacked non-linear
functions to extract hidden patterns, making them good candidates for modelling complex likelihood surfaces and multimodal posterior distributions. 

To structure the DNN, one first constructs an input layer of $D$ neurons corresponding to the $D$ values of $q^\m$.   This is followed by a number of
hidden layers with differing numbers of neurons that attempt to model the data, which ultimately feed into an output layer of $D$ neurons containing the $D$ gradients of the 
log-likelihood.  In general the number of neurons in the hidden layers are greater than the number of neurons in the input and output layers.  In this study, we use fully connected
(or dense) layers where each neuron is connected to every other neuron in both the preceeding and successive layers.  The output of each neuron in layer $k$ is given by
\begin{equation}\label{eq:activation_function}
    a^k_j = \phi \left( \sum_{i=1}^{\,\,n_{(k-1)}} w^k_{ij}  a^{k-1}_i + b^k_j \right),
 \end{equation}
where $a^{k-1}_i$ are the input values coming from layer $k-1$, $w^k_{ij}$ is a matrix of weights defining the strength of connectivity between neurons, $b^k_j$ is a bias vector which ensures activation
of the neurons even if the inputs are zero, $n_{(k-1)}$ is the number of neurons in the $k-1$ layer, and $\phi$ is an activation function which introduces non-linearity into the network.


\subsection{Tuning the DNN}
Each artificial network has a number of hyperparameters that need to be fine tuned for optimum performance.  In general these hyperparameters can be split into two sets: those that define
the network structure (e.g. the number of hidden layers and neurons, the dropout rate, the choice of activation function etc.) and those that define the network training (e.g. the learning rate, 
batch size, number of training epochs etc.).   We will deal with each set in order below.


\subsubsection{Tuning the network structure}
At the beginning of the project, we used the \emph{Keras} (v2.3.1) library~{\cite{chollet2015keras}, a widely used deep learning API running on top of the machine learning platform \emph{TensorFlow} (v1.14.0)~\cite{tensorflow2015-whitepaper},  to build and train our DNN.  To stack the dense layers of the network we use the \texttt{Sequential} model within
\emph{Keras}.  After some initial testing, we decided on having three hidden layers in the network: two small and one large.  To choose the number of neurons in each 
layer we use the \texttt{RandomSearch} function which randomly picks several possible DNN architectures amongst given layer sizes, trains each of them for a given number
of epochs and returns the DNN with the lowest error based on our test set.   While there is essentially no limit to the number of layers and neurons per layer, we wanted
to keep the network training duration below 1 hour, so that it does not present a new computational bottleneck.  Secondly, the more neurons
in the network, the more time it takes to make new predictions, which is a potential issue given that we need to predict gradient values sequentially along the Hamiltonian
trajectory.

\begin{figure}[t]
       \includegraphics[height=10.5cm, width=0.45\textwidth]{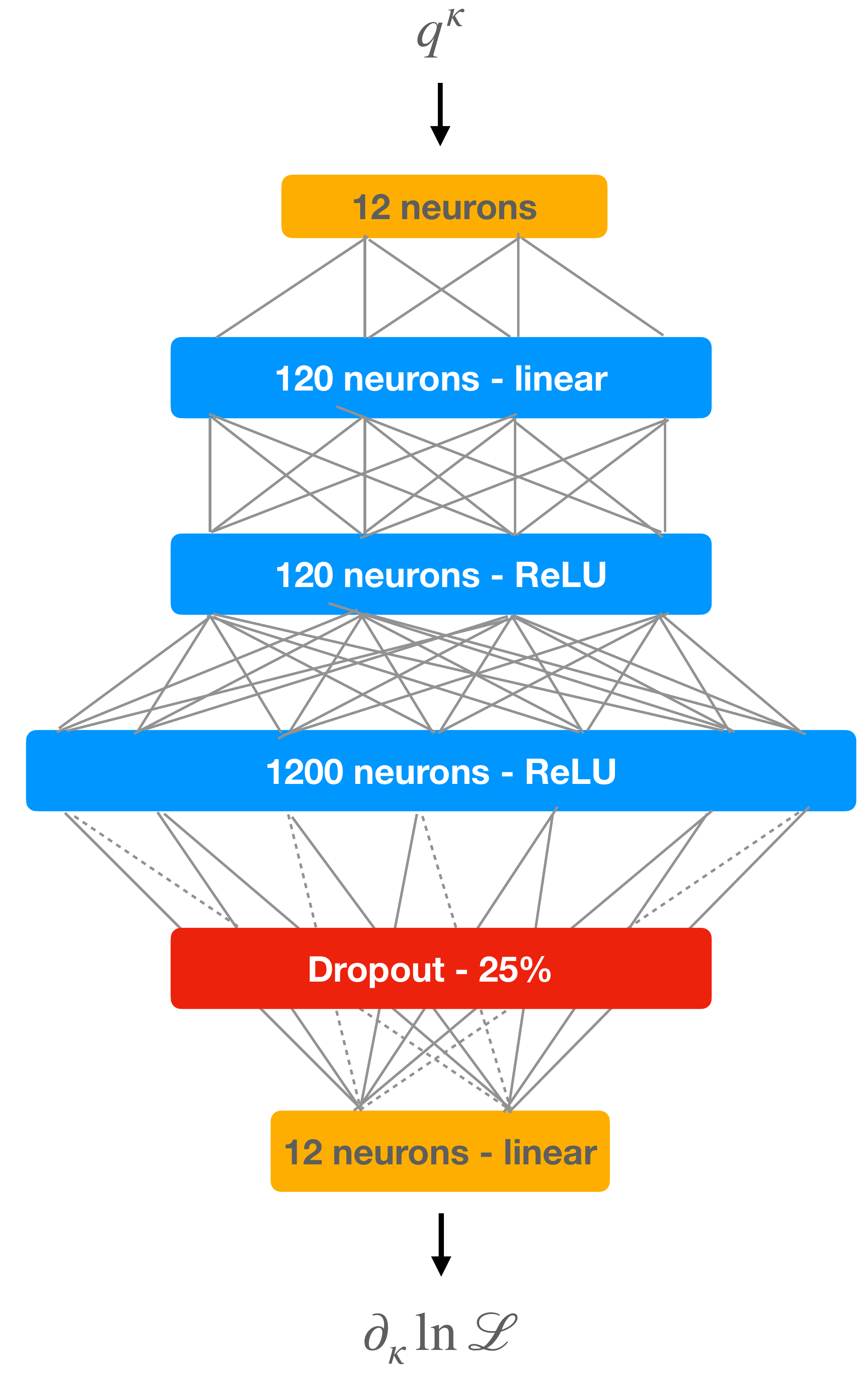}
        \caption{The DNN architecture for approximating the $D$ gradients of the log-likelihood given the $D$-dimensional coordinate position.  This architecture consists of three fully connected hidden internal layers and a single 25\% dropout layer.  A linear activation function is used for the first hidden layer, while the common ReLU function is used to introduce non-linearity into the remaining hidden layers.}
        \label{fig:dnnarch}
\end{figure}

To initialize \texttt{RandomSearch}, we allowed the small layers to have 
between $(D, 5D, 10D)$ neurons, with the large layer being composed of $(50D, 100D, 200D)$ neurons.  In the end \texttt{RandomSearch} returned an optimal
architecture of $(10D, 100D, 10D)$ neurons.  However, a quick manual test showed that an architecture of $(10D, 10D, 100D)$ neurons actually gave the same performance while
reducing the training time by a factor of about 1.4.  This is due to the fact that when the largest layer is connected to the output layer, there are fewer weights to update in the 
overall network.  A final test further showed that we achieved an even better performance if we included a dropout layer of 25\% between the largest layer and the output layer.  This
dropout layer is a common regularization technique that prevents overfitting by the network~\cite{srivastava_dropout_2014}.

Once the global architecture of the network was set, the next step was to choose the activation function appearing in Eq.~(\ref{eq:activation_function}).  A common function in DNNs is the 
rectified linear unit (ReLU) function $\phi = max\{0,x\}$ which allows us to introduce non-linearity into the DNN.  In the beginning we applied this function to all internal layers, but after some testing we found a better performance when applying a linear function to the first hidden layer, and the ReLU function for the other two.  Finally, we decided to initialize the weights and biases in the network by drawing from a normal kernel initializer.  We provide a graphical representation of the DNN architecture in Fig.~\ref{fig:dnnarch}


\subsubsection{Tuning the network training}
After constructing the DNN, the next step was to optimize the network training.    At this step of the project, and mostly motivated by a change of hardward to ARM-based Apple silicon processors,
we switched from TensorFlow to PyTorch (v2.2.2)    Fine tuning the weights and biases is done through an iterative training procedure on the training set, which is 
split into batches of equal batchsize, each picked randomly but left disjoint such that every example is used and used once. Starting from the initial state of weights and biases, we use a mean 
squared error loss function (MSE) between the predicted and numerical gradients to measure how incorrect the tuning is on the batch considered~\cite{10.5555/3378999}, i.e.
\begin{equation}
      MSE = \frac{1}{BD}\sum_{i=0}^{B-1}\sum_{\k=0}^{D-1}\left[\left(\dk\lnL\right)_{\num}^i - \left(\dk\lnL\right)_{DNN}^i\right]^2,
      \label{eq:mse}
\end{equation}
where $B$ is the batchsize.  By computing the gradient of the loss function with respect to all weights and biases, we can update them in their inverse gradient direction with a learning rate which
controls the step size between old and new values. This operation is repeated until every batch is treated, which completes an epoch.  And the process is then repeated for several epochs allowing the 
network to treat the training samples in a different order.    To properly solve Eqn.~(\ref{eq:mse}), we first shuffle and normalize the Phase I data set.  By shuffling the data, we avoid bias in the data and
improve generalization.   This procedure also breaks any patterns that exist in the sequential data that the DNN might learn instead of the underlying global model.  Normalizing the data is important, as
if one dimension in the network has a variance orders of magnitude larger than others, it will dominate the loss function and prevent the network from 
learning in other dimensions.   To avoid this possibility, it is common to standardize the training set by centering distributions in each positional and gradient dimension around zero and scaling them to unit variance.
For this, we use scikit-learn's \texttt{StandardScaler}~\cite{scikit-learn}. We then split the Phase I data into 
three parts: training data (70\%), validation data (20\%) and test data (10\%).  Finally, we need to tune three DNN hyperparameters: (a) the learning rate schedule (b) the batch size and (c) the number
of training epochs.

Initially, we implemented the stochastic decent method using the Adam optimizer with the learning rate set to its default value of $R_L = 10^{-3}$~\cite{2014arXiv1412.6980K} and with a fixed the number of epochs of 100.  While this
static learning rate worked well, we found a better performance using the PyTorch  \texttt{ExponentialLR} function.  This function, defined as
\beq
R_L^{e+1} = \gamma R_L^e,
\eeq
divides the learning rate at each epoch $e$ by a factor $\gamma$.  This causes the learning rate to decrease abruptly in the initial epochs and slows later on.  As the learning rate aspires to zero, without ever reaching it, an
excessively small learning rate can have a negative impact on the learning performance.  To avoid this, we being with a learning rate of $R_L = 10^{-3}$, and decrease exponentially to a lower bound of $R_L = 10^{-4}$, using
$\gamma = 0.99$.

\begin{figure*}
\begin{center}
\begin{minipage}[b]{.49\linewidth}
\centering
\epsfig{file=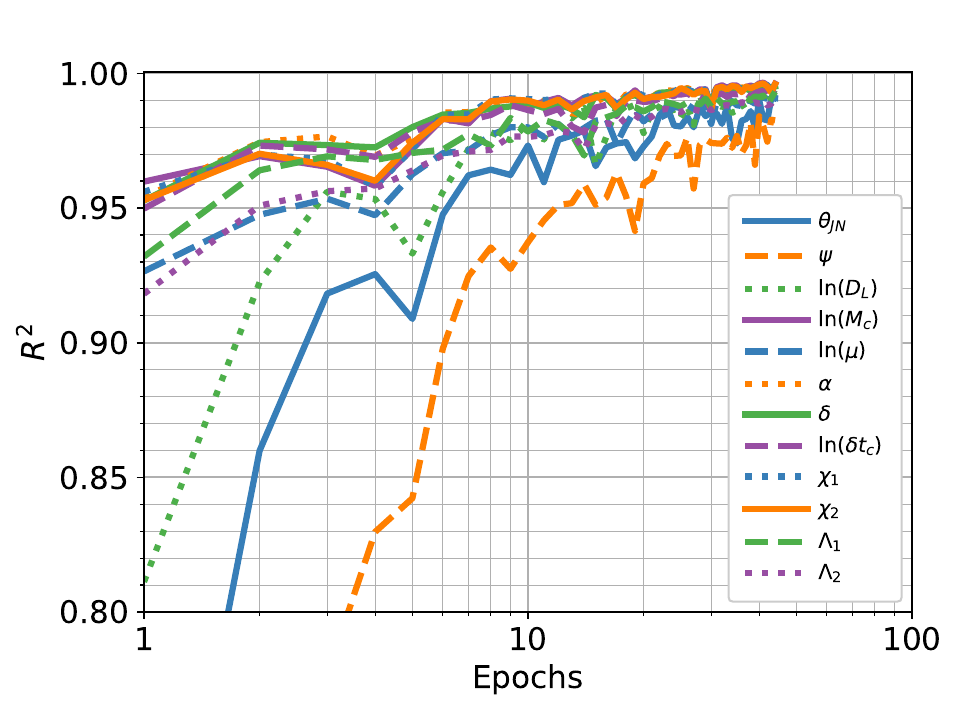, width=\textwidth, height=2.7in}
\end{minipage} \hfill
\begin{minipage}[b]{.49\linewidth}
\centering
\epsfig{file=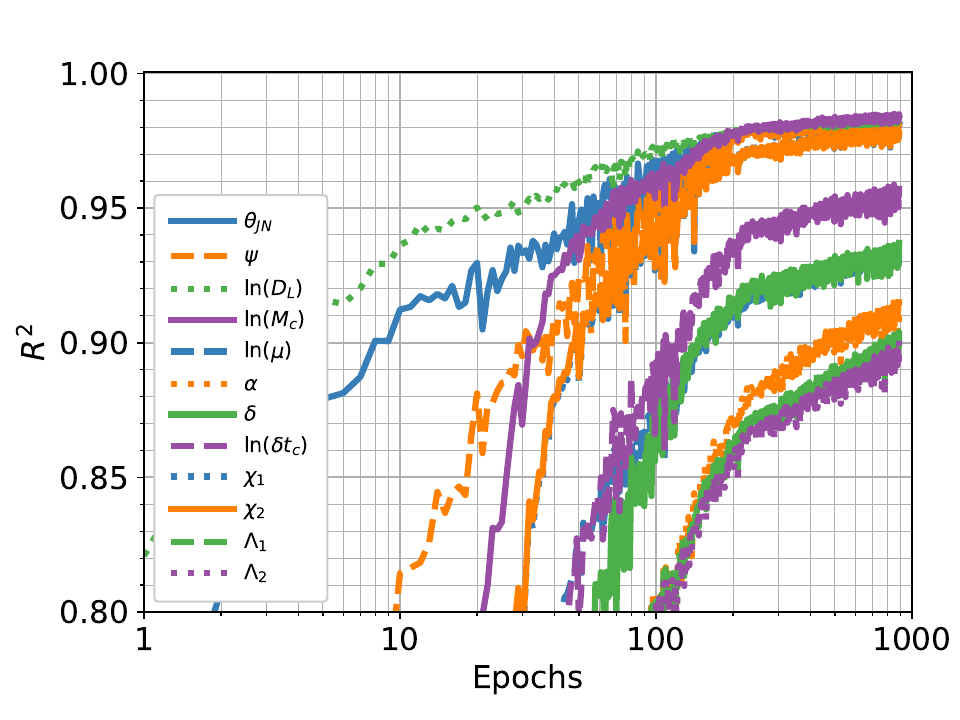, width=\textwidth, height=2.7in}
\end{minipage}
 \caption{Evolution of the coefficient of determination, $R^2$, as a function of the number of training epochs.  We observe that the DNN converges much faster for a high-SNR unimodal source like GW170817 as opposed
 to a low-SNR mulitmodal source such as GW190425.}
\label{fig:R2_evolution}
\end{center}
\end{figure*}

The next hyper-parameter to tune is the batch size $B$.  While there is a general recommendation that the batch size be kept small~\cite{2018arXiv180407612M}, we decided to test our network on twelve different batch sizes
from $B\in[2^4, 2^{15}]$ using the same training and test sets as used previously.  In general, small batch sizes ensure fast training, while large batch sizes ensure more precise estimation of
the stochastic gradient decent.  We found that training on batch sizes between $B\in[16,128]$ quickly reached a low error loss, indicating that the network has good generalization abilities.  Conversely,
larger batch sizes displayed a much noisier behaviour, with the error curves either monotonically decreasing or failing to converge, indicating poor generalization.  Amongst the performing batch sizes, we 
eventually chose a batch size of 128 as it produced a gradient decent as good as a batch size of 32, but at the same time reduced the runtime for 60 epochs from 21 min to 12 min.

The final hyper-parameter to tune is the number of training epochs.  Our initial studies began with fixed epoch numbers for both GW170817 and GW190425.  To evaluate the performance of the training model, we 
 calculate the coefficient of determination, $R^2$, defined by
\bea
R^2 & = & 1-\frac{SS_{res}}{SS_{tot}} \\ 
       & = & 1 - \frac{\sum_{i=1}^N \left[\left(\dk\lnL\right)_i^{num} - \left(\dk\lnL\right)_i^{DNN} \right]^2}{\sum_{i=1}^N \left[\left(\dk\lnL\right)_i^{num} - \left(\overline{\dk\lnL}\right)^{num} \right]^2},\nonumber
\eea
where the superscripts $num$ and $DNN$ denote numerical and DNN gradients respectively, $N$ is the number of data points, $SS_{res}$ is the residual sum of squares, $SS_{tot}$ is the total sum of squares, and
\beq
\left(\overline{\dk\lnL}\right)^{num} = \frac{1}{N}\sum_{i=1}^N \left(\dk\lnL\right)_i^{num}
\eeq
is the gradient mean, using the test data set.  If the approximate data is equivalent to the numerical data, then $SS_{res} = 0$ and $R^2=1$, and the distribution lies along a diagonal line.  The $R^2$ value has the common interpretation that,
for example, a value of $R^2 = 0.9$ implies that 90\% of the data fits the regression model.  There is no hard rule on choosing the value of $R^2$, but clearly, the higher the $R^2$ score, the better the model fit for the gradients.

Beginning with GW170817, we initially set the number of training epochs to 1000 and a threshold of $R^2\geq 0.9$.  In all of our preliminary investigations, we found that for GW170817-like events, we achieved $R^2$-scores of almost
unity in a small number of training epochs.  We also found that the $R^2$ score for the polarization angle $\psi$ has no effect on the Phase III performance.  This lead us to set the criteria of $R^2 \geq 0.99$ (for all parameters besides $\psi$) or 500 training epochs for high-SNR events with a unimodal distribution in the sky position.  On the left hand side of Fig.~\ref{fig:R2_evolution}, we plot
  $R^2$ evolution curves as a function of epoch number for an optimal training run.  In this case, we reach the criterion of $R^2\geq 0.99$ in less than 50 epochs, needing a Phase II runtime of just nine minutes.  In one run, the training
  did take the full 500 epochs to reach the threshold of $R^2>0.99$, and in this case, the DNN training took 100 minutes.   In Fig.~\ref{fig:GW170817_R2} (Appendix~\ref{sec:R2_scores}), we plot the $R^2$ scores at the end of a training run, using the 
  test data set.  We can see that all distributions lie along the diagonal.

For GW190425, the situation was more complex.  Starting with the same configuration as GW170817, we failed to reach a high $R^2$ value.  Our first test was to increase the amount of Phase I data.  Given bimodal distributions across the range of the priors
in many of the parameters, this made no real improvement to the model fit, but increased the Phase I runtime to the order of a day.  We then decided to lower the $R^2$ threshold to $R^2\geq 0.95$, kept the number of accepted Phase I trajectories at 3300 and
experimented in increasing the number of training epochs.  In each case, we ran a restricted Phase III of the algorithm where we monitored the number of SISs acquired and the runtime.  We eventually found that it was possible to reach the threshold
of $R^2\geq 0.95$, but we required 20,000 training epochs, and a training runtime of almost a week.  We also found that even in this case, we produced only a small number of extra SISs.  Given our investigations, we then took the decision to set the criteria of
$R^2 \geq 0.9$ or 1000 training epochs for signals having a multimodal sky distribution. 

On the right hand side of Fig.~\ref{fig:R2_evolution}, we plot
  $R^2$ evolution curves as a function of epoch number for the slowest training run we observed.  In this case, we reach the criterion of $R^2\geq 0.9$ in just under 900 epochs, needing a Phase II runtime of 200 minutes.  In the best run we observed, the training
  took around 400 epochs to reach the threshold of $R^2>0.9$, with a DNN training time of 80 minutes.   In Fig.~\ref{fig:GW190425_R2} (Appendix~\ref{sec:R2_scores}), we plot the $R^2$ scores at the end of a training run, using the 
  test data set.  We can see that while all distributions lie along the diagonal, they are more diffuse than the GW170817 case.  We also noticed, in all runs, that while many of the parameters hugely exceeded the $R^2$-threshold, the tidal deformation parameters, $\Lambda_i$, were consistently the slowest parameters to 
  reach the $R^2$ threshold.  This is not surprising as these are higher order corrections to the gravitational waveform, and are not very well resolved for low-SNR sources.

To see the benefit of using the DNN gradients, we looked at the runtime for a single 200 step trajectory using the 128 seconds of GW170817 data.  A naive approach, where the $12D$ gradients of the log-likelihood ratio are calculated using the full likelihood calculation (Eqn~\ref{eq:likelihood_ratio}), took 7 minutes to produce a single trajectory.  Swapping out the full likelihood ratio with a calculation based on relative binning, allowed us to generate a single trajectory in 1.8 seconds, a speed-up of approximately 200.    By then replacing the 
relative binning likelihood gradients with DNN gradients, a single trajectory took 0.06 seconds to generate.  This is a factor of 30 times faster than the relative binning gradient based trajectories, and a factor of 7000 times faster than a naive calculation.



\subsection{Final strucutre of \dhmc}
In this section,  we detail the final structure of the algorithm :
    \begin{enumerate}
     \item Phase I: run $3\,300$  trajectories in parallel using numerical gradients of fixed length with $l=200$ steps and with a stepsize $\epsilon = 5\times10^{-3}$. Each core uses the same trigger point, but a different random seed for the momenta, thus ensuring diverging trajectories.  For every trajectory, record each visited position $q^\mu$ and the associated $D$-dimensional numerical gradients $(\partial_\mu\lnL)_{\num}$ at that point.
      \item Phase II:
        \begin{itemize}
          \item Train the DNN for on $\{q^{\k},  (\partial_\k\lnL)_{\num}\}$ from Phase I with a batchsize of 128, using the stopping criteria of $R^2>0.99$ or 500 training epochs for unimodal sources in sky position, or $R^2>0.9$ or 1000 training epochs for sources with a bimodal structure in sky position.
          \item Recompute the leapfrog scales using the covariance matrix of the Phase I chain points.  For each parameter we take the $n$
samples gathered in Phase I and construct the $n\times D$ data matrix 
$Q^{\k}_{n} = \left[q^{\k}_1, ..., q^{\k}_n\right]$.  From this, we calculate the $D\times D$ covariance matrix
\beq
C_{\k\s} = \frac{1}{n-1}\left[\left(Q^{\k}_{n} -\bar{q}^{\k}\right)^{T} \left(Q^{\s}_{n} -\bar{q}^{\s}\right)\right],
\eeq
where $\bar{q}^{\k} = \sum_i q^{\k}_{i} / n$  is a $D$-dimensional vector containing the means for each parameter.   At the beginning of Phase III we then set the scales to $s^{\k}=\s_{\k} = (C_{\k\k})^{1/2}$.
        \end{itemize}
        \item Phase III: run $N$ trajectories using the DNN approximation for the gradients, with $\e\in{\mathcal N}[5\times10^{-3},1.5\times10^{-3}]$ (such that $\e^{\k} = s^{\k}\e$) and $l\in\mathcal{U}(50, 400)$.
        \begin{itemize}
          \item Recompute the leapfrog scales very $10^4$ trajectories, stopping when there is less that $10\%$ change in numerical values             
          \item Every 50,000 trajectories calculate the integrated autocorrelation time and the number of SISs already acquired.  If we have acquired more than 90\% of the required SISs, half the number of trajecotries and adjust $N$ appropriately to achieve the required number of SISs
          \item If we have 100 consecutive rejections, then we run a short ($l=50$) trajectory using numerical gradients.
        \end{itemize}
    \end{enumerate}

\begin{figure*}
\begin{center}
\begin{minipage}[b]{.49\linewidth}
\centering
\epsfig{file=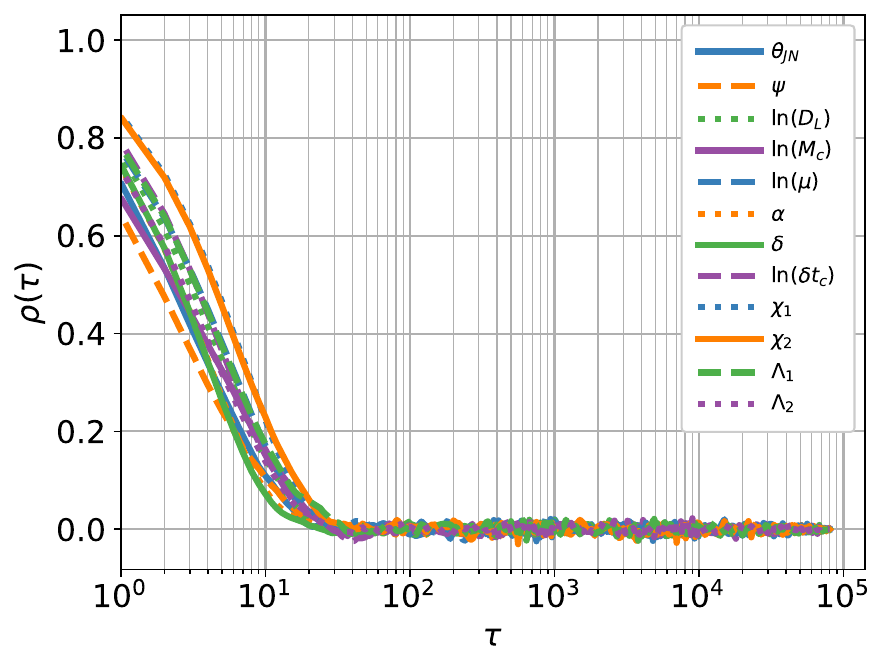, width=0.95\textwidth, height=2.5in}
\end{minipage} \hfill
\begin{minipage}[b]{.49\linewidth}
\centering
\epsfig{file=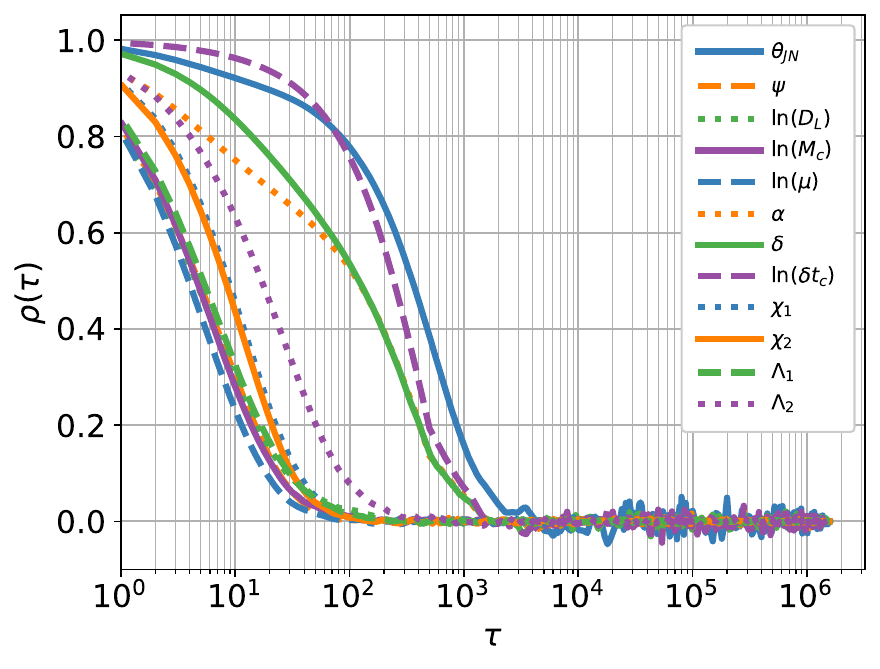, width=0.95\textwidth, height=2.5in}
\end{minipage}
 \caption{Autocorrelation functions $\rho(\tau)$ against lag $\tau$ for GW170817 (left) and GW190425 (right).  In the case of GW170817, the autocorrelation function generally crosses zero in less than 100 lags, while in the case of GW190425 it usually crosses zero in under $10^4$ lags.}
\label{fig:autocorrelations}
\end{center}
\end{figure*}


\section{Application to GW170817 and GW190425}
In this final section we apply \dhmc to the full 128 second public data set for GW170817~\cite{gwosc170817} and GW190425~\cite{gwosc190425}.  To validate the algorithm, we compare the posterior density samples from \dhmc with the public LVK samples produced using LALInferenceMCMC.     Our final runs were conducted using  Macbook Pro laptops with a 10-core 3.2 GHz M1/M2 processors and 16 GB of LPDDR5 memory.  For each case, we conducted multiple runs.  We present the results for each source below.


\subsection{GW170817}
For GW170817, Phase I took approximately 20 minutes to produce 3300 accepted trajectories, giving 660,000 data points, with an average acceptance rate of $\geq90\%$.  The Phase II DNN training took between 10-100 minutes, or 40-500 epochs, to run.  The wide
range in training time comes from the fact that PyTorch randomizes the weights in Eqn~(\ref{eq:activation_function}) which affects the convergence of the DNN training.  Our goal in this study was to obtain 5000 SISs to compare with the LVK public results.  
Consequently, Phase III took between 90-120 minutes to obtain the 5000 SISs, with an acceptance rate of $\geq80\%$.  On the left hand side of Fig.~\ref{fig:autocorrelations}, we plot the autocorrelation function for the Phase III chains.  We can see that all chains drop to zero at a similar rate.  In all runs, the slowest autocorrelation
crosses zero in a few tens of lags, giving an integrated autocorrelation time of $\tau_{int}(N)  \sim10$, meaning that the algorithm produced 1 SIS/sec for this source.  The total runtime for the GW170817 inference was between 2-4 hours.

In Fig.~\ref{fig:GW170817_posterior} we present a comparison of the posterior densities from both the LVK public samples (blue) and the \dhmc samples (orange), based on 5000 SISs.  We observe an excellent agreement between the two distributions.  We should note that the LVK did not provide posterior distributions for the polarization angle $\psi$ or the time of coalescence $t_c$, and fixed the sky position based on EM information regarding the host galaxy.  However, we assumed that each of these were free parameters for the inference, and present posterior distributions accordingly.  In columns two and three of Table~\ref{tab:hmc_results} we present a quantative comparison of the medians and 90\% credible intervals based on the posterior distributions.  In all cases, we again see excellent agreement between the LVK and \dhmc results.

On an interesting note, the final LVK sky area for GW170817 was 28 deg$^2$, with a luminosity distance estimate of $39.9_{+7}^{-15}$ Mpcs.  At the end of Phase I (which again took around 20 minutes on 10 cores), our algorithm predicts a sky area of 15 deg$^2$ and a luminosity distance of $37.6^{+8}_{-16}$ Mpcs.  This suggests that by increasing the number of cores used, \dhmc also has the capability of predicting accurate 3D error boxes in low latency.


\subsection{GW190425}
For GW190425, Phase I took approximately 25 minutes to produce 3300 accepted trajectories, giving 660,000 data points, with an average acceptance rate of $\geq80\%$.   As we know from the trigger that this event had a bimodal sky distribution, we started
five cores at the trigger value, and five cores at a flipped sky position given by
\bea
\a &=& \a + \pi \nonumber\\
\delta &=& -\delta
\eea
This allows us to explore both sky modes and acquire information.  The Phase II DNN training took between 80-200 minutes, or 400-900 epochs, to run, again with the wide
range in training time comes from the fact that PyTorch randomizes the weights in Eqn~(\ref{eq:activation_function}) which affects the convergence of the DNN training.  For GW190425,  LVK public data contained 53,900 SISs.   To make a comparison, we 
ran \dhmc three times in parallel, producing 17,500 SISs from the combined runs, which we compared the first 17,500 samples from the LVK data.   Consequently, each of the Phase III runs took between 90-120 minutes, with an acceptance rate of $\geq80\%$.  On the right hand side of Fig.~\ref{fig:autocorrelations}, we plot the autocorrelation function for the Phase III chains.  In contrast with the unimodal case of GW170817, we can see that the chains with the longest autocorrelations are those that are highly bimodal, i.e.
$\{\ctjn, \a, \delta, \ln(\delta t_c)\}$.  In all runs, the slowest autocorrelation
crosses zero in a few thousands of lags, giving an integrated autocorrelation time of $\tau_{int}(N)  \in[200,300]$, meaning that the algorithm produced 1 SIS every 26 secs for this source.  The total runtime for the GW190425 inference was around 2.5 days.

In Fig.~\ref{fig:GW190425_posterior} we present a comparison of the posterior densities from both the LVK public samples (blue) and the \dhmc samples (orange), based on 17,500 SISs.  We observe an excellent agreement between the two distributions for the intrinsic parameters.  Looking at Fig.~\ref{fig:skymaps}, we can see that the posterior distributions cover a large swathe of the sky.  This lead to a mismatch in certain parameters such as the inclination angle $\theta_{JN}$, however, it demonstrates that the algorithm is capable of exploring multimodal structures on the likelihood surface.  In columns four and five of Table~\ref{tab:hmc_results} we present a quantative comparison of the medians and 90\% credible intervals based on the posterior distributions.  In all cases, we again see excellent agreement between the LVK and \dhmc results for the intrinsic parameters.

\begin{figure*}[th]
       \includegraphics[width=0.9\textwidth, height=8.59cm]{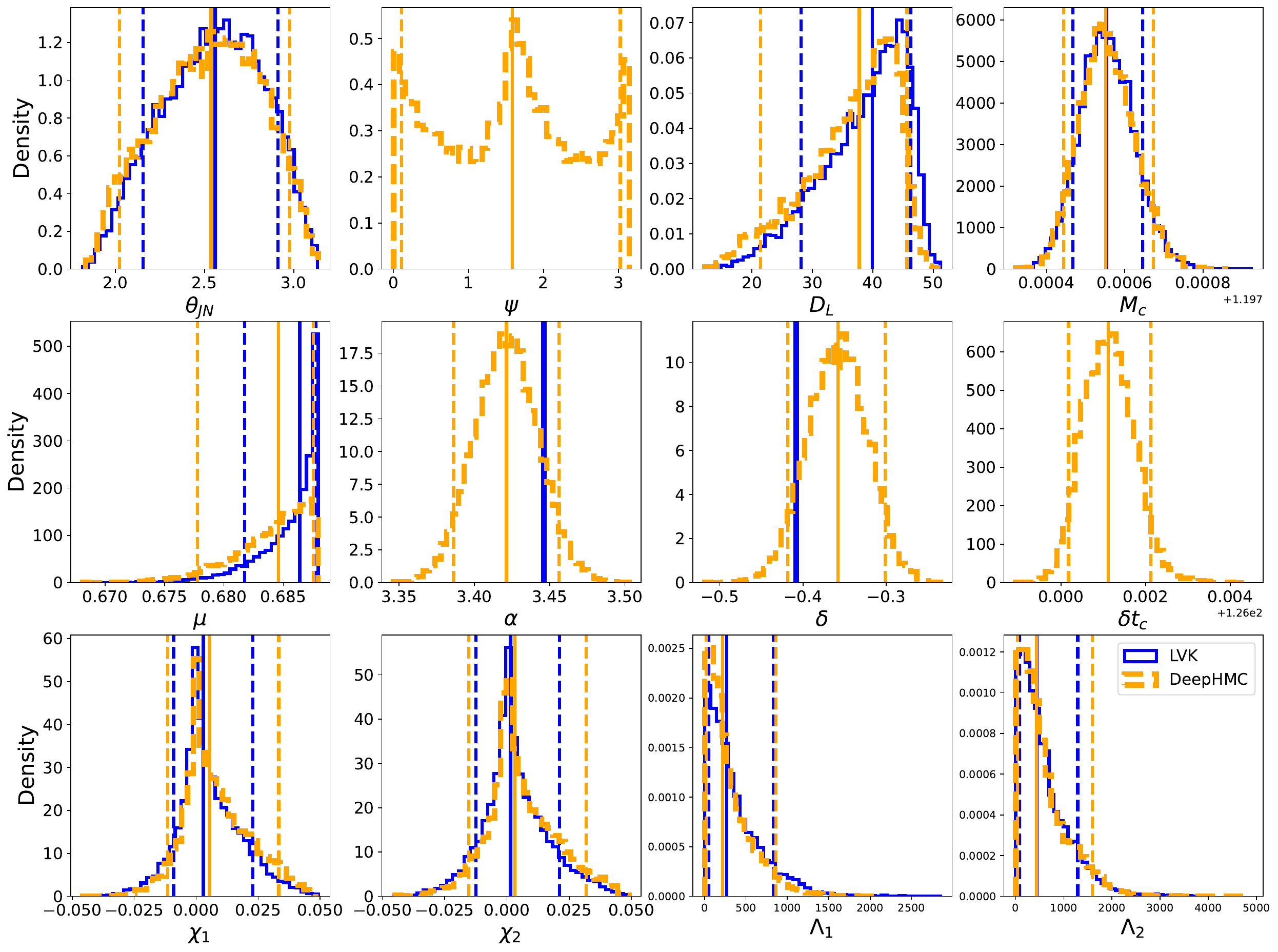}
        \caption{A comparison between the LVK public data (blue) and \dhmc (orange) posterior distributions based on 5000 SISs.  The LVK samples do not contain the polarization angle $\psi$ or time to coalescence $t_c$, and are based on a fixed sky position associated with the host galaxy (blue solid lines for $(\a,\delta)$).}
        \label{fig:GW170817_posterior}.   
\end{figure*}

\begin{figure*}[th]
       \includegraphics[width=0.9\textwidth, height=8.59cm]{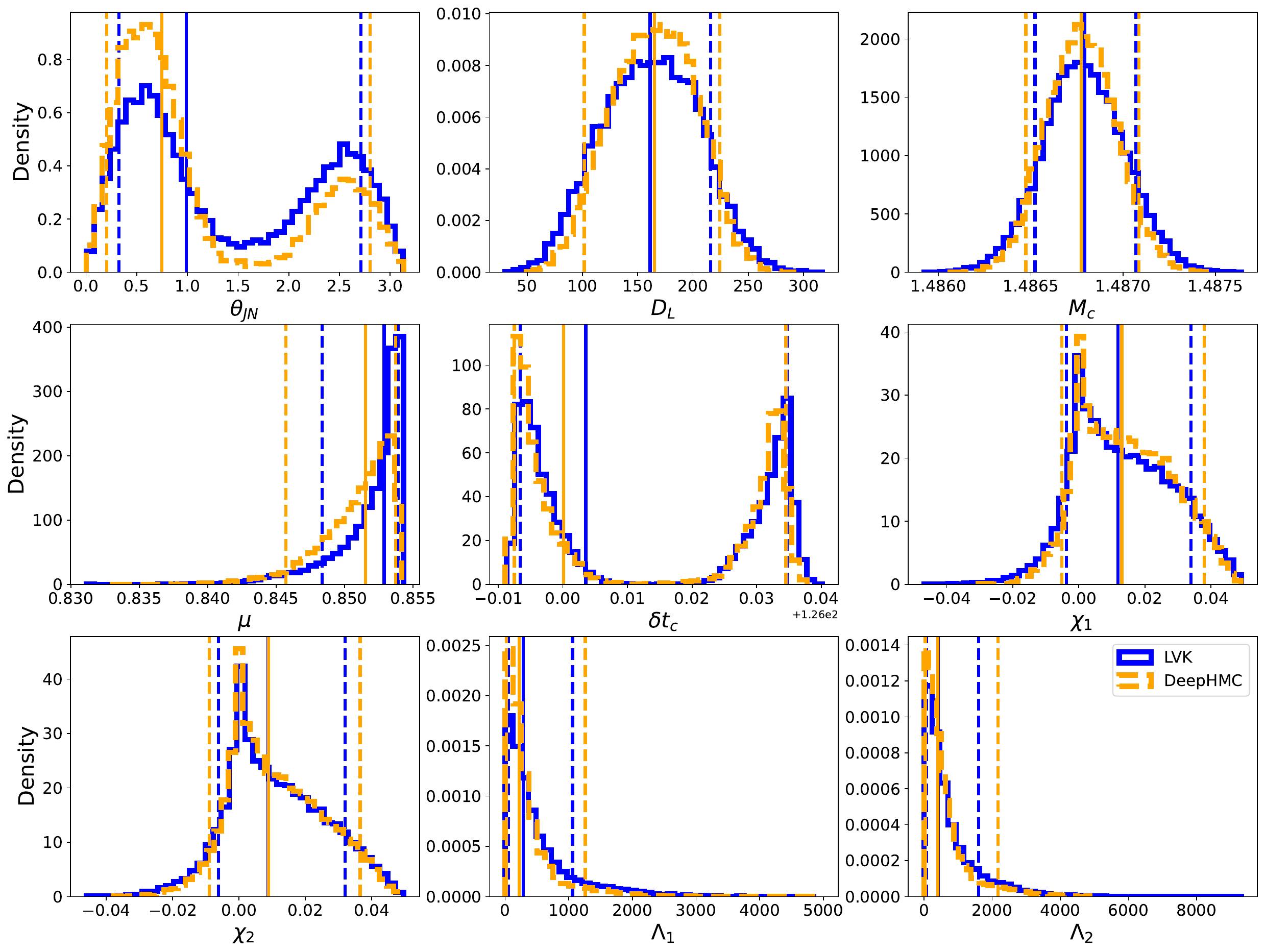}
        \caption{A comparison between the LVK public data (blue) and \dhmc (orange) posterior distributions based on 17,500 SISs.  The LVK samples do not contain the polarization angle $\psi$.}
        \label{fig:GW190425_posterior}.   
\end{figure*}

\begin{figure*}
\begin{center}
\begin{minipage}[b]{.49\linewidth}
\centering
\epsfig{file=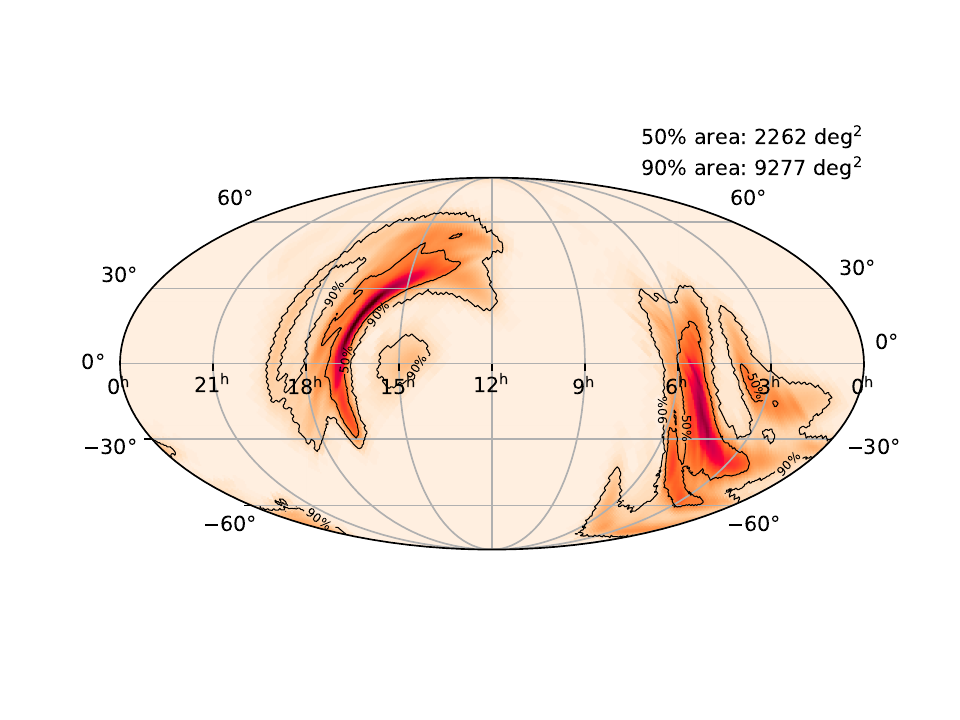, width=0.95\textwidth, height=2.5in}
\end{minipage} \hfill
\begin{minipage}[b]{.49\linewidth}
\centering
\epsfig{file=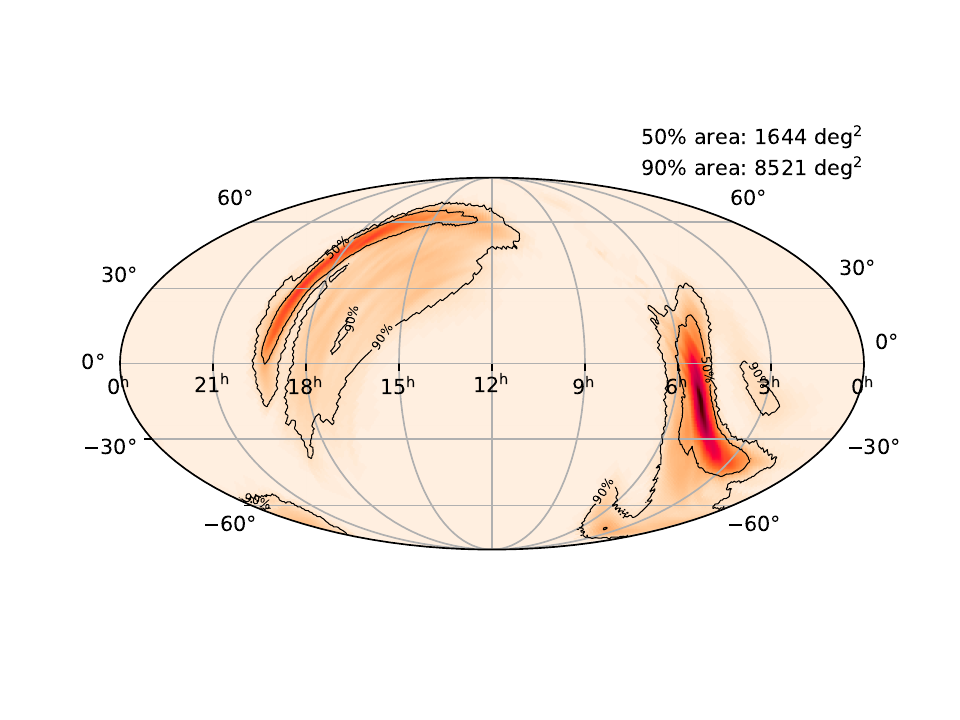, width=0.95\textwidth, height=2.5in}
\end{minipage}
 \caption{Sky maps for both the LVK and \dhmc posterior distributions.}
\label{fig:skymaps}
\end{center}
\end{figure*}


\begin{table*}
\centering
  \begin{tabular}{lcccc}
    \hline
    \multicolumn{5}{c}{\,\,\,\,\,\,\,\,\,\,\,\,\,\,\,\,\,\,\,\,\,\,\,\,\,\,\,GW170817\,\,\,\,\,\,\,\,\,\,\,\,\,\,\,\,\,\,\,\,\,\,\,\,\,\,\,\,\,\,\,\,\,\,\,\,\,\,\,\,\,\,\,\,\,\,\,\,\,\,\,\,\,\,\,\,\,\,\,\,\,\,\,\,\,\,\,\,\,\,\,\,\,\,\,\,\,\,\,GW190425}\\
    \hline
    Parameter & LVK & \dhmc & LVK & \dhmc\\
    \hline
    $\theta_{JN}$/rad & ${2.560}_{-0.492}^{+0.418}$ & ${2.552}_{-0.512}^{+0.424}$  & --- & ---\\
    $D_L$/Mpc & ${39.956}_{-15.017}^{+7.343}$ & ${38.322}_{-16.038}^{+7.317}$ & ${162}_{-72}^{+70}$ & ${165}_{-63}^{+59}$\\
    $m_1/\msun$ & ${1.479}_{-0.094}^{+0.155}$ & ${1.508}_{-0.092}^{+0.115}$ & ${1.81}_{-0.09}^{+0.18}$ & ${1.86}_{-0.11}^{+0.14}$\\
    $m_2/\msun$ & ${1.281}_{-0.116}^{+0.086}$ & ${1.257}_{-0.085}^{+0.080}$ & ${1.62}_{-0.14}^{+0.08}$ & ${1.57}_{-0.11}^{+0.10}$\\
    $\mathcal{M}/\msun$ & ${1.1976}_{-0.0001}^{+0.0001}$ & ${1.1975}_{-0.0001}^{+0.0001}$ & ${1.48678}_{-0.00035}^{+0.00037}$ & ${1.48677}_{-0.00030}^{+0.00031}$\\
    $q$ & ${0.866}_{-0.153}^{+0.121}$ & ${0.833}_{-0.112}^{+0.111}$  & ${0.894}_{-0.154}^{+0.096}$ & ${0.846}_{-0.114}^{+0.106}$\\
    $\alpha$/rad & ${3.446}_{-0.0}^{+0.0}$ & ${3.421}_{-0.034}^{+0.037}$ &--- & ---\\
    $\delta$/rad & ${-0.408}_{-0.0}^{+0.0}$ & ${-0.357}_{-0.062}^{+0.056}$ & --- & ---\\
    $t_c$/s & ${1187008882.4312}_{-0.0}^{+0.0}$ & ${1187008882.4312}_{-0.0009}^{+0.0011}$ & ${1240215503.01}_{-0.01}^{+0.03}$ & ${1240215503.01}_{-0.01}^{+0.03}$\\
    $\chi_1$ & ${0.003}_{-0.017}^{+0.026}$ & ${0.004}_{-0.016}^{+0.024}$ & ${0.012}_{-0.020}^{+0.027}$ & ${0.013}_{-0.018}^{+0.025}$\\
    $\chi_2$ & ${0.001}_{-0.019}^{+0.027}$ & ${0.002}_{-0.020}^{+0.024}$ & ${0.0090}_{-0.020}^{+0.029}$ & ${0.0089}_{-0.018}^{+0.028}$\\
    $\chi_{eff}$ & ${0.003}_{-0.008}^{+0.015}$ & ${0.004}_{-0.008}^{+0.012}$ & ${0.0118}_{-0.0130}^{+0.0153}$ & ${0.0126}_{-0.0125}^{+0.0139}$ \\
    $\Lambda_1$ & ${269}_{-243.124}^{+776.100}$ & ${284.706}_{-255.367}^{+753.335}$ & ${320}_{-295}^{+1440}$ & ${222}_{-204}^{+1036}$\\
    $\Lambda_2$ & ${445.771}_{-406.928}^{+1145.381}$ & ${396.792}_{-361.217}^{+1121.821}$  & ${483}_{-443}^{+2050}$ & ${405}_{-370}^{+1770}$\\
    $\tilde{\Lambda}$ & ${534.159}_{-347.852}^{+717.791}$ & ${551.131}_{-378.726}^{+719.500}$  & ${444}_{-325}^{+1134}$ & ${316}_{-219}^{+1048}$\\ \\
     \hline
  \end{tabular}
  \caption{Comparison of the median and $90\%$ credible intervals for the posterior distributions produced  by the LVK and \dhmc for GW170817 and GW190425 respectively.  We note that the constant LVK values for the sky position and time of coalescence for GW170817 are based on the associated EM counterpart and the detection timestamp respectively.  For GW190425, we present results for unimodal posterior distributions only.}
  \label{tab:hmc_results}
\end{table*}




\section{Conclusion and next steps}
In this article we have presented a deep neural network based HMC algorithm, called \dhmc, which can be applied to the inference of binary neutron star coalescences in GW astronomy.  This algorithm, fully written in Python, is particularly suited to cases where the waveform lengths are quite long, and inference of both spins and tidal deformations are required.  This version is a major leap forward over previous versions of our algorithm as it dispenses with the need for any a priori knowledge of the modality of the target density when approximating the gradients of the target density.  

To demonstrate the validity of the algorithm, we compared its performance against the public LVK results for the BNS mergers GW170817 and GW190425.  In both cases, we were able to reproduce posterior distributions for the binary parameters that demonstrated a high level of fidelity with the public results.  We were also able to demonstrate in an apples-to-apples comparison that \dhmc is very efficient in comparison to state of the art algorithms in terms of both runtime and the cost of producing a SIS.  The inference took 2-4 hours for GW170817 and around 2.5 days for GW190425, with a cost of 1 SIS every second for GW170817, and 1 SIS every 26 seconds for GW190425.

In a future work, we plan to apply the algorithm to NSBH and sub-solar mass systems.


\begin{acknowledgments}
The authors would like to thank the internal LVK referees for their useful comments.

This research has made use of data or software obtained from the Gravitational Wave Open Science Center (gwosc.org), a service of the LIGO Scientific Collaboration, the Virgo Collaboration, and KAGRA. This material is based upon work supported by NSF's LIGO Laboratory which is a major facility fully funded by the National Science Foundation, as well as the Science and Technology Facilities Council (STFC) of the United Kingdom, the Max-Planck-Society (MPS), and the State of Niedersachsen/Germany for support of the construction of Advanced LIGO and construction and operation of the GEO600 detector. Additional support for Advanced LIGO was provided by the Australian Research Council. Virgo is funded, through the European Gravitational Observatory (EGO), by the French Centre National de Recherche Scientifique (CNRS), the Italian Istituto Nazionale di Fisica Nucleare (INFN) and the Dutch Nikhef, with contributions by institutions from Belgium, Germany, Greece, Hungary, Ireland, Japan, Monaco, Poland, Portugal, Spain. KAGRA is supported by Ministry of Education, Culture, Sports, Science and Technology (MEXT), Japan Society for the Promotion of Science (JSPS) in Japan; National Research Foundation (NRF) and Ministry of Science and ICT (MSIT) in Korea; Academia Sinica (AS) and National Science and Technology Council (NSTC) in Taiwan.

\end{acknowledgments}

\bibliography{deepHMC.bib}

\begin{thebibliography}{62}
\expandafter\ifx\csname natexlab\endcsname\relax\def\natexlab#1{#1}\fi
\expandafter\ifx\csname bibnamefont\endcsname\relax
  \def\bibnamefont#1{#1}\fi
\expandafter\ifx\csname bibfnamefont\endcsname\relax
  \def\bibfnamefont#1{#1}\fi
\expandafter\ifx\csname citenamefont\endcsname\relax
  \def\citenamefont#1{#1}\fi
\expandafter\ifx\csname url\endcsname\relax
  \def\url#1{\texttt{#1}}\fi
\expandafter\ifx\csname urlprefix\endcsname\relax\def\urlprefix{URL }\fi
\providecommand{\bibinfo}[2]{#2}
\providecommand{\eprint}[2][]{\url{#2}}

\bibitem[{\citenamefont{Abbott et~al.}(2016)}]{Abbott:2016blz}
\bibinfo{author}{\bibfnamefont{B.~P.} \bibnamefont{Abbott}}
  \bibnamefont{et~al.} (\bibinfo{collaboration}{LIGO Scientific Collaboration,
  Virgo Collaboration}), \bibinfo{journal}{Phys. Rev. Lett.}
  \textbf{\bibinfo{volume}{116}}, \bibinfo{pages}{061102}
  (\bibinfo{year}{2016}), \eprint{1602.03837}.

\bibitem[{\citenamefont{Abbott et~al.}(2019)}]{LIGOScientific:2018mvr}
\bibinfo{author}{\bibfnamefont{B.~P.} \bibnamefont{Abbott}}
  \bibnamefont{et~al.} (\bibinfo{collaboration}{LIGO Scientific Collaboration,
  Virgo Collaboration}), \bibinfo{journal}{Phys. Rev. X}
  \textbf{\bibinfo{volume}{9}}, \bibinfo{pages}{031040} (\bibinfo{year}{2019}),
  \eprint{1811.12907}.

\bibitem[{\citenamefont{Abbott
  et~al.}(2021{\natexlab{a}})}]{LIGOScientific:2020ibl}
\bibinfo{author}{\bibfnamefont{R.}~\bibnamefont{Abbott}} \bibnamefont{et~al.}
  (\bibinfo{collaboration}{LIGO Scientific, Virgo}), \bibinfo{journal}{Phys.
  Rev. X} \textbf{\bibinfo{volume}{11}}, \bibinfo{pages}{021053}
  (\bibinfo{year}{2021}{\natexlab{a}}), \eprint{2010.14527}.

\bibitem[{\citenamefont{Abbott
  et~al.}(2021{\natexlab{b}})}]{LIGOScientific:2021djp}
\bibinfo{author}{\bibfnamefont{R.}~\bibnamefont{Abbott}} \bibnamefont{et~al.}
  (\bibinfo{collaboration}{LIGO Scientific, VIRGO, KAGRA})
  (\bibinfo{year}{2021}{\natexlab{b}}), \eprint{2111.03606}.

\bibitem[{\citenamefont{Abbott et~al.}(2017)}]{TheLIGOScientific:2017qsa}
\bibinfo{author}{\bibfnamefont{B.~P.} \bibnamefont{Abbott}}
  \bibnamefont{et~al.} (\bibinfo{collaboration}{LIGO Scientific Collaboration,
  Virgo Collaboration}), \bibinfo{journal}{Phys. Rev. Lett.}
  \textbf{\bibinfo{volume}{119}}, \bibinfo{pages}{161101}
  (\bibinfo{year}{2017}), \eprint{1710.05832}.

\bibitem[{\citenamefont{Abbott et~al.}(2020)}]{LIGOScientific:2020aai}
\bibinfo{author}{\bibfnamefont{B.~P.} \bibnamefont{Abbott}}
  \bibnamefont{et~al.} (\bibinfo{collaboration}{LIGO Scientific, Virgo}),
  \bibinfo{journal}{Astrophys. J. Lett.} \textbf{\bibinfo{volume}{892}},
  \bibinfo{pages}{L3} (\bibinfo{year}{2020}), \eprint{2001.01761}.

\bibitem[{\citenamefont{Abbott
  et~al.}(2021{\natexlab{c}})}]{LIGOScientific:2021qlt}
\bibinfo{author}{\bibfnamefont{R.}~\bibnamefont{Abbott}} \bibnamefont{et~al.}
  (\bibinfo{collaboration}{LIGO Scientific, KAGRA, VIRGO}),
  \bibinfo{journal}{Astrophys. J. Lett.} \textbf{\bibinfo{volume}{915}},
  \bibinfo{pages}{L5} (\bibinfo{year}{2021}{\natexlab{c}}),
  \eprint{2106.15163}.

\bibitem[{\citenamefont{Hastings}(1970)}]{hastings_1970}
\bibinfo{author}{\bibfnamefont{W.~K.} \bibnamefont{Hastings}},
  \bibinfo{journal}{Biometrika} \textbf{\bibinfo{volume}{57}},
  \bibinfo{pages}{97} (\bibinfo{year}{1970}).

\bibitem[{\citenamefont{{Metropolis} et~al.}(1953)\citenamefont{{Metropolis},
  {Rosenbluth}, {Rosenbluth}, {Teller}, and {Teller}}}]{metropolis_1953}
\bibinfo{author}{\bibfnamefont{N.}~\bibnamefont{{Metropolis}}},
  \bibinfo{author}{\bibfnamefont{A.~W.} \bibnamefont{{Rosenbluth}}},
  \bibinfo{author}{\bibfnamefont{M.~N.} \bibnamefont{{Rosenbluth}}},
  \bibinfo{author}{\bibfnamefont{A.~H.} \bibnamefont{{Teller}}},
  \bibnamefont{and} \bibinfo{author}{\bibfnamefont{E.}~\bibnamefont{{Teller}}},
  \bibinfo{journal}{J. Chem. Phys.} \textbf{\bibinfo{volume}{21}},
  \bibinfo{pages}{1087} (\bibinfo{year}{1953}).

\bibitem[{\citenamefont{Skilling}(2006)}]{skilling_2006}
\bibinfo{author}{\bibfnamefont{J.}~\bibnamefont{Skilling}},
  \bibinfo{journal}{Bayesian Anal.} \textbf{\bibinfo{volume}{1}},
  \bibinfo{pages}{833} (\bibinfo{year}{2006}).

\bibitem[{\citenamefont{Ashton et~al.}(2019)}]{Ashton:2018jfp}
\bibinfo{author}{\bibfnamefont{G.}~\bibnamefont{Ashton}} \bibnamefont{et~al.},
  \bibinfo{journal}{Astrophys. J. Suppl.} \textbf{\bibinfo{volume}{241}},
  \bibinfo{pages}{27} (\bibinfo{year}{2019}), \eprint{1811.02042}.

\bibitem[{\citenamefont{Speagle}(2020)}]{Speagle_2020}
\bibinfo{author}{\bibfnamefont{J.~S.} \bibnamefont{Speagle}},
  \bibinfo{journal}{Monthly Notices of the Royal Astronomical Society}
  \textbf{\bibinfo{volume}{493}}, \bibinfo{pages}{3132–3158}
  (\bibinfo{year}{2020}), ISSN \bibinfo{issn}{1365-2966},
  \urlprefix\url{http://dx.doi.org/10.1093/mnras/staa278}.

\bibitem[{\citenamefont{{Neal}}(2011)}]{2011hmcm.book..113N}
\bibinfo{author}{\bibfnamefont{R.}~\bibnamefont{{Neal}}}, in
  \emph{\bibinfo{booktitle}{Handbook of Markov Chain Monte Carlo}}
  (\bibinfo{publisher}{CRC press}, \bibinfo{year}{2011}), pp.
  \bibinfo{pages}{113--162}.

\bibitem[{\citenamefont{Punturo et~al.}(2010)\citenamefont{Punturo, Abernathy,
  Acernese, Allen, Andersson, Arun, Barone, Barr, Barsuglia, Beker
  et~al.}}]{Punturo_2010}
\bibinfo{author}{\bibfnamefont{M.}~\bibnamefont{Punturo}},
  \bibinfo{author}{\bibfnamefont{M.}~\bibnamefont{Abernathy}},
  \bibinfo{author}{\bibfnamefont{F.}~\bibnamefont{Acernese}},
  \bibinfo{author}{\bibfnamefont{B.}~\bibnamefont{Allen}},
  \bibinfo{author}{\bibfnamefont{N.}~\bibnamefont{Andersson}},
  \bibinfo{author}{\bibfnamefont{K.}~\bibnamefont{Arun}},
  \bibinfo{author}{\bibfnamefont{F.}~\bibnamefont{Barone}},
  \bibinfo{author}{\bibfnamefont{B.}~\bibnamefont{Barr}},
  \bibinfo{author}{\bibfnamefont{M.}~\bibnamefont{Barsuglia}},
  \bibinfo{author}{\bibfnamefont{M.}~\bibnamefont{Beker}},
  \bibnamefont{et~al.}, \bibinfo{journal}{Classical and Quantum Gravity}
  \textbf{\bibinfo{volume}{27}}, \bibinfo{pages}{194002}
  (\bibinfo{year}{2010}),
  \urlprefix\url{https://dx.doi.org/10.1088/0264-9381/27/19/194002}.

\bibitem[{\citenamefont{Abac et~al.}(2025)\citenamefont{Abac, Abramo, Albanesi,
  Albertini, Agapito, Agathos, Albertus, Andersson, Andrade, Andreoni
  et~al.}}]{abac2025scienceeinsteintelescope}
\bibinfo{author}{\bibfnamefont{A.}~\bibnamefont{Abac}},
  \bibinfo{author}{\bibfnamefont{R.}~\bibnamefont{Abramo}},
  \bibinfo{author}{\bibfnamefont{S.}~\bibnamefont{Albanesi}},
  \bibinfo{author}{\bibfnamefont{A.}~\bibnamefont{Albertini}},
  \bibinfo{author}{\bibfnamefont{A.}~\bibnamefont{Agapito}},
  \bibinfo{author}{\bibfnamefont{M.}~\bibnamefont{Agathos}},
  \bibinfo{author}{\bibfnamefont{C.}~\bibnamefont{Albertus}},
  \bibinfo{author}{\bibfnamefont{N.}~\bibnamefont{Andersson}},
  \bibinfo{author}{\bibfnamefont{T.}~\bibnamefont{Andrade}},
  \bibinfo{author}{\bibfnamefont{I.}~\bibnamefont{Andreoni}},
  \bibnamefont{et~al.}, \emph{\bibinfo{title}{The science of the einstein
  telescope}} (\bibinfo{year}{2025}), \eprint{2503.12263},
  \urlprefix\url{https://arxiv.org/abs/2503.12263}.

\bibitem[{\citenamefont{Reitze et~al.}(2019)\citenamefont{Reitze, Adhikari,
  Ballmer, Barish, Barsotti, Billingsley, Brown, Chen, Coyne, Eisenstein
  et~al.}}]{reitze2019cosmicexploreruscontribution}
\bibinfo{author}{\bibfnamefont{D.}~\bibnamefont{Reitze}},
  \bibinfo{author}{\bibfnamefont{R.~X.} \bibnamefont{Adhikari}},
  \bibinfo{author}{\bibfnamefont{S.}~\bibnamefont{Ballmer}},
  \bibinfo{author}{\bibfnamefont{B.}~\bibnamefont{Barish}},
  \bibinfo{author}{\bibfnamefont{L.}~\bibnamefont{Barsotti}},
  \bibinfo{author}{\bibfnamefont{G.}~\bibnamefont{Billingsley}},
  \bibinfo{author}{\bibfnamefont{D.~A.} \bibnamefont{Brown}},
  \bibinfo{author}{\bibfnamefont{Y.}~\bibnamefont{Chen}},
  \bibinfo{author}{\bibfnamefont{D.}~\bibnamefont{Coyne}},
  \bibinfo{author}{\bibfnamefont{R.}~\bibnamefont{Eisenstein}},
  \bibnamefont{et~al.}, \emph{\bibinfo{title}{Cosmic explorer: The u.s.
  contribution to gravitational-wave astronomy beyond ligo}}
  (\bibinfo{year}{2019}), \eprint{1907.04833},
  \urlprefix\url{https://arxiv.org/abs/1907.04833}.

\bibitem[{\citenamefont{Evans et~al.}(2023)\citenamefont{Evans, Corsi, Afle,
  Ananyeva, Arun, Ballmer, Bandopadhyay, Barsotti, Baryakhtar, Berger
  et~al.}}]{evans2023cosmicexplorersubmissionnsf}
\bibinfo{author}{\bibfnamefont{M.}~\bibnamefont{Evans}},
  \bibinfo{author}{\bibfnamefont{A.}~\bibnamefont{Corsi}},
  \bibinfo{author}{\bibfnamefont{C.}~\bibnamefont{Afle}},
  \bibinfo{author}{\bibfnamefont{A.}~\bibnamefont{Ananyeva}},
  \bibinfo{author}{\bibfnamefont{K.~G.} \bibnamefont{Arun}},
  \bibinfo{author}{\bibfnamefont{S.}~\bibnamefont{Ballmer}},
  \bibinfo{author}{\bibfnamefont{A.}~\bibnamefont{Bandopadhyay}},
  \bibinfo{author}{\bibfnamefont{L.}~\bibnamefont{Barsotti}},
  \bibinfo{author}{\bibfnamefont{M.}~\bibnamefont{Baryakhtar}},
  \bibinfo{author}{\bibfnamefont{E.}~\bibnamefont{Berger}},
  \bibnamefont{et~al.}, \emph{\bibinfo{title}{Cosmic explorer: A submission to
  the nsf mpsac nggw subcommittee}} (\bibinfo{year}{2023}),
  \eprint{2306.13745}, \urlprefix\url{https://arxiv.org/abs/2306.13745}.

\bibitem[{\citenamefont{Hu and
  Veitch}(2025)}]{hu2025costsbayesianparameterestimation}
\bibinfo{author}{\bibfnamefont{Q.}~\bibnamefont{Hu}} \bibnamefont{and}
  \bibinfo{author}{\bibfnamefont{J.}~\bibnamefont{Veitch}},
  \emph{\bibinfo{title}{Costs of bayesian parameter estimation in
  third-generation gravitational wave detectors: a review of acceleration
  methods}} (\bibinfo{year}{2025}), \eprint{2412.02651},
  \urlprefix\url{https://arxiv.org/abs/2412.02651}.

\bibitem[{\citenamefont{Cornish}(2013)}]{cornish2013fastfishermatriceslazy}
\bibinfo{author}{\bibfnamefont{N.~J.} \bibnamefont{Cornish}},
  \emph{\bibinfo{title}{Fast fisher matrices and lazy likelihoods}}
  (\bibinfo{year}{2013}), \eprint{1007.4820},
  \urlprefix\url{https://arxiv.org/abs/1007.4820}.

\bibitem[{\citenamefont{{Zackay} et~al.}(2018)\citenamefont{{Zackay}, {Dai},
  and {Venumadhav}}}]{2018arXiv180608792Z}
\bibinfo{author}{\bibfnamefont{B.}~\bibnamefont{{Zackay}}},
  \bibinfo{author}{\bibfnamefont{L.}~\bibnamefont{{Dai}}}, \bibnamefont{and}
  \bibinfo{author}{\bibfnamefont{T.}~\bibnamefont{{Venumadhav}}},
  \bibinfo{journal}{arXiv e-prints} \bibinfo{eid}{arXiv:1806.08792}
  (\bibinfo{year}{2018}), \eprint{1806.08792}.

\bibitem[{\citenamefont{Morisaki}(2021)}]{Morisaki_2021}
\bibinfo{author}{\bibfnamefont{S.}~\bibnamefont{Morisaki}},
  \bibinfo{journal}{Physical Review D} \textbf{\bibinfo{volume}{104}}
  (\bibinfo{year}{2021}), ISSN \bibinfo{issn}{2470-0029},
  \urlprefix\url{http://dx.doi.org/10.1103/PhysRevD.104.044062}.

\bibitem[{\citenamefont{Canizares et~al.}(2015)\citenamefont{Canizares, Field,
  Gair, Raymond, Smith, and Tiglio}}]{Canizares_2015}
\bibinfo{author}{\bibfnamefont{P.}~\bibnamefont{Canizares}},
  \bibinfo{author}{\bibfnamefont{S.~E.} \bibnamefont{Field}},
  \bibinfo{author}{\bibfnamefont{J.}~\bibnamefont{Gair}},
  \bibinfo{author}{\bibfnamefont{V.}~\bibnamefont{Raymond}},
  \bibinfo{author}{\bibfnamefont{R.}~\bibnamefont{Smith}}, \bibnamefont{and}
  \bibinfo{author}{\bibfnamefont{M.}~\bibnamefont{Tiglio}},
  \bibinfo{journal}{Physical Review Letters} \textbf{\bibinfo{volume}{114}}
  (\bibinfo{year}{2015}), ISSN \bibinfo{issn}{1079-7114},
  \urlprefix\url{http://dx.doi.org/10.1103/PhysRevLett.114.071104}.

\bibitem[{\citenamefont{Smith et~al.}(2016)\citenamefont{Smith, Field,
  Blackburn, Haster, Pürrer, Raymond, and Schmidt}}]{Smith_2016}
\bibinfo{author}{\bibfnamefont{R.}~\bibnamefont{Smith}},
  \bibinfo{author}{\bibfnamefont{S.~E.} \bibnamefont{Field}},
  \bibinfo{author}{\bibfnamefont{K.}~\bibnamefont{Blackburn}},
  \bibinfo{author}{\bibfnamefont{C.-J.} \bibnamefont{Haster}},
  \bibinfo{author}{\bibfnamefont{M.}~\bibnamefont{Pürrer}},
  \bibinfo{author}{\bibfnamefont{V.}~\bibnamefont{Raymond}}, \bibnamefont{and}
  \bibinfo{author}{\bibfnamefont{P.}~\bibnamefont{Schmidt}},
  \bibinfo{journal}{Physical Review D} \textbf{\bibinfo{volume}{94}}
  (\bibinfo{year}{2016}), ISSN \bibinfo{issn}{2470-0029},
  \urlprefix\url{http://dx.doi.org/10.1103/PhysRevD.94.044031}.

\bibitem[{\citenamefont{Lange et~al.}(2018{\natexlab{a}})\citenamefont{Lange,
  O'Shaughnessy, and Rizzo}}]{lange2018rapidaccurateparameterinference}
\bibinfo{author}{\bibfnamefont{J.}~\bibnamefont{Lange}},
  \bibinfo{author}{\bibfnamefont{R.}~\bibnamefont{O'Shaughnessy}},
  \bibnamefont{and} \bibinfo{author}{\bibfnamefont{M.}~\bibnamefont{Rizzo}},
  \emph{\bibinfo{title}{Rapid and accurate parameter inference for coalescing,
  precessing compact binaries}} (\bibinfo{year}{2018}{\natexlab{a}}),
  \eprint{1805.10457}, \urlprefix\url{https://arxiv.org/abs/1805.10457}.

\bibitem[{\citenamefont{Morisaki et~al.}(2023)\citenamefont{Morisaki, Smith,
  Tsukada, Sachdev, Stevenson, Talbot, and
  Zimmerman}}]{morisaki2023rapidlocalizationinferencecompact}
\bibinfo{author}{\bibfnamefont{S.}~\bibnamefont{Morisaki}},
  \bibinfo{author}{\bibfnamefont{R.}~\bibnamefont{Smith}},
  \bibinfo{author}{\bibfnamefont{L.}~\bibnamefont{Tsukada}},
  \bibinfo{author}{\bibfnamefont{S.}~\bibnamefont{Sachdev}},
  \bibinfo{author}{\bibfnamefont{S.}~\bibnamefont{Stevenson}},
  \bibinfo{author}{\bibfnamefont{C.}~\bibnamefont{Talbot}}, \bibnamefont{and}
  \bibinfo{author}{\bibfnamefont{A.}~\bibnamefont{Zimmerman}},
  \emph{\bibinfo{title}{Rapid localization and inference on compact binary
  coalescences with the advanced ligo-virgo-kagra gravitational-wave detector
  network}} (\bibinfo{year}{2023}), \eprint{2307.13380},
  \urlprefix\url{https://arxiv.org/abs/2307.13380}.

\bibitem[{\citenamefont{Roulet et~al.}(2024)\citenamefont{Roulet, Mushkin,
  Wadekar, Venumadhav, Zackay, and Zaldarriaga}}]{Roulet_2024}
\bibinfo{author}{\bibfnamefont{J.}~\bibnamefont{Roulet}},
  \bibinfo{author}{\bibfnamefont{J.}~\bibnamefont{Mushkin}},
  \bibinfo{author}{\bibfnamefont{D.}~\bibnamefont{Wadekar}},
  \bibinfo{author}{\bibfnamefont{T.}~\bibnamefont{Venumadhav}},
  \bibinfo{author}{\bibfnamefont{B.}~\bibnamefont{Zackay}}, \bibnamefont{and}
  \bibinfo{author}{\bibfnamefont{M.}~\bibnamefont{Zaldarriaga}},
  \bibinfo{journal}{Physical Review D} \textbf{\bibinfo{volume}{110}}
  (\bibinfo{year}{2024}), ISSN \bibinfo{issn}{2470-0029},
  \urlprefix\url{http://dx.doi.org/10.1103/PhysRevD.110.044010}.

\bibitem[{\citenamefont{Williams et~al.}(2021)\citenamefont{Williams, Veitch,
  and Messenger}}]{Williams_2021}
\bibinfo{author}{\bibfnamefont{M.~J.} \bibnamefont{Williams}},
  \bibinfo{author}{\bibfnamefont{J.}~\bibnamefont{Veitch}}, \bibnamefont{and}
  \bibinfo{author}{\bibfnamefont{C.}~\bibnamefont{Messenger}},
  \bibinfo{journal}{Physical Review D} \textbf{\bibinfo{volume}{103}}
  (\bibinfo{year}{2021}), ISSN \bibinfo{issn}{2470-0029},
  \urlprefix\url{http://dx.doi.org/10.1103/PhysRevD.103.103006}.

\bibitem[{\citenamefont{Dax et~al.}(2025)\citenamefont{Dax, Green, Gair, Gupte,
  Pürrer, Raymond, Wildberger, Macke, Buonanno, and Schölkopf}}]{Dax_2025}
\bibinfo{author}{\bibfnamefont{M.}~\bibnamefont{Dax}},
  \bibinfo{author}{\bibfnamefont{S.~R.} \bibnamefont{Green}},
  \bibinfo{author}{\bibfnamefont{J.}~\bibnamefont{Gair}},
  \bibinfo{author}{\bibfnamefont{N.}~\bibnamefont{Gupte}},
  \bibinfo{author}{\bibfnamefont{M.}~\bibnamefont{Pürrer}},
  \bibinfo{author}{\bibfnamefont{V.}~\bibnamefont{Raymond}},
  \bibinfo{author}{\bibfnamefont{J.}~\bibnamefont{Wildberger}},
  \bibinfo{author}{\bibfnamefont{J.~H.} \bibnamefont{Macke}},
  \bibinfo{author}{\bibfnamefont{A.}~\bibnamefont{Buonanno}}, \bibnamefont{and}
  \bibinfo{author}{\bibfnamefont{B.}~\bibnamefont{Schölkopf}},
  \bibinfo{journal}{Nature} \textbf{\bibinfo{volume}{639}},
  \bibinfo{pages}{49–53} (\bibinfo{year}{2025}), ISSN
  \bibinfo{issn}{1476-4687},
  \urlprefix\url{http://dx.doi.org/10.1038/s41586-025-08593-z}.

\bibitem[{\citenamefont{Porter and Carr\'e}(2014)}]{Porter:2013wwa}
\bibinfo{author}{\bibfnamefont{E.~K.} \bibnamefont{Porter}} \bibnamefont{and}
  \bibinfo{author}{\bibfnamefont{J.}~\bibnamefont{Carr\'e}},
  \bibinfo{journal}{Class. Quant. Grav.} \textbf{\bibinfo{volume}{31}},
  \bibinfo{pages}{145004} (\bibinfo{year}{2014}), \eprint{1311.7539}.

\bibitem[{\citenamefont{Bouffanais and Porter}(2019)}]{Bouffanais:2018hoz}
\bibinfo{author}{\bibfnamefont{Y.}~\bibnamefont{Bouffanais}} \bibnamefont{and}
  \bibinfo{author}{\bibfnamefont{E.~K.} \bibnamefont{Porter}},
  \bibinfo{journal}{Phys. Rev. D} \textbf{\bibinfo{volume}{100}},
  \bibinfo{pages}{104023} (\bibinfo{year}{2019}), \eprint{1810.07443}.

\bibitem[{\citenamefont{Nitz}(2024)}]{nitz2024robustrapidsimplegravitationalwave}
\bibinfo{author}{\bibfnamefont{A.~H.} \bibnamefont{Nitz}},
  \emph{\bibinfo{title}{Robust, rapid, and simple gravitational-wave parameter
  estimation}} (\bibinfo{year}{2024}), \eprint{2410.05190},
  \urlprefix\url{https://arxiv.org/abs/2410.05190}.

\bibitem[{\citenamefont{Wong et~al.}(2023)\citenamefont{Wong, Isi, and
  Edwards}}]{wong2023fastgravitationalwaveparameter}
\bibinfo{author}{\bibfnamefont{K.~W.~K.} \bibnamefont{Wong}},
  \bibinfo{author}{\bibfnamefont{M.}~\bibnamefont{Isi}}, \bibnamefont{and}
  \bibinfo{author}{\bibfnamefont{T.~D.~P.} \bibnamefont{Edwards}},
  \emph{\bibinfo{title}{Fast gravitational wave parameter estimation without
  compromises}} (\bibinfo{year}{2023}), \eprint{2302.05333},
  \urlprefix\url{https://arxiv.org/abs/2302.05333}.

\bibitem[{\citenamefont{Duane et~al.}(1987)\citenamefont{Duane, Kennedy,
  Pendleton, and Roweth}}]{duane_1987}
\bibinfo{author}{\bibfnamefont{S.}~\bibnamefont{Duane}},
  \bibinfo{author}{\bibfnamefont{A.~D.} \bibnamefont{Kennedy}},
  \bibinfo{author}{\bibfnamefont{B.}~\bibnamefont{Pendleton}},
  \bibnamefont{and} \bibinfo{author}{\bibfnamefont{D.}~\bibnamefont{Roweth}},
  \bibinfo{journal}{Phys. Lett. B.} \textbf{\bibinfo{volume}{195}},
  \bibinfo{pages}{216} (\bibinfo{year}{1987}).

\bibitem[{\citenamefont{Hajian}(2007)}]{Hajian:2006mt}
\bibinfo{author}{\bibfnamefont{A.}~\bibnamefont{Hajian}},
  \bibinfo{journal}{Phys. Rev. D} \textbf{\bibinfo{volume}{75}},
  \bibinfo{pages}{083525} (\bibinfo{year}{2007}), \eprint{astro-ph/0608679}.

\bibitem[{\citenamefont{Buonanno et~al.}(2009)\citenamefont{Buonanno, Iyer,
  Ochsner, Pan, and Sathyaprakash}}]{Buonanno_2009}
\bibinfo{author}{\bibfnamefont{A.}~\bibnamefont{Buonanno}},
  \bibinfo{author}{\bibfnamefont{B.~R.} \bibnamefont{Iyer}},
  \bibinfo{author}{\bibfnamefont{E.}~\bibnamefont{Ochsner}},
  \bibinfo{author}{\bibfnamefont{Y.}~\bibnamefont{Pan}}, \bibnamefont{and}
  \bibinfo{author}{\bibfnamefont{B.~S.} \bibnamefont{Sathyaprakash}},
  \bibinfo{journal}{Physical Review D} \textbf{\bibinfo{volume}{80}}
  (\bibinfo{year}{2009}).

\bibitem[{\citenamefont{GWOSC}(2017)}]{gwosc170817}
\bibinfo{author}{\bibnamefont{GWOSC}} (\bibinfo{year}{2017}),
  \bibinfo{note}{\url{https://www.gw-openscience.org/eventapi/html/GWTC-1-confident/GW170817/v3}}.

\bibitem[{\citenamefont{GWOSC}(2019)}]{gwosc190425}
\bibinfo{author}{\bibnamefont{GWOSC}} (\bibinfo{year}{2019}),
  \bibinfo{note}{\url{https://gwosc.org/eventapi/html/GWTC-2.1-confident/GW190425/v3/}}.

\bibitem[{\citenamefont{Finn}(1992)}]{Finn:1992wt}
\bibinfo{author}{\bibfnamefont{L.~S.} \bibnamefont{Finn}},
  \bibinfo{journal}{Phys. Rev. D} \textbf{\bibinfo{volume}{46}},
  \bibinfo{pages}{5236} (\bibinfo{year}{1992}), \eprint{gr-qc/9209010}.

\bibitem[{\citenamefont{Owen}(1996)}]{Owen:1995tm}
\bibinfo{author}{\bibfnamefont{B.~J.} \bibnamefont{Owen}},
  \bibinfo{journal}{Phys. Rev. D} \textbf{\bibinfo{volume}{53}},
  \bibinfo{pages}{6749} (\bibinfo{year}{1996}), \eprint{gr-qc/9511032}.

\bibitem[{\citenamefont{Cornish and Porter}(2006)}]{Cornish:2006ry}
\bibinfo{author}{\bibfnamefont{N.~J.} \bibnamefont{Cornish}} \bibnamefont{and}
  \bibinfo{author}{\bibfnamefont{E.~K.} \bibnamefont{Porter}},
  \bibinfo{journal}{Class. Quant. Grav.} \textbf{\bibinfo{volume}{23}},
  \bibinfo{pages}{S761} (\bibinfo{year}{2006}), \eprint{gr-qc/0605085}.

\bibitem[{\citenamefont{Cornish and Porter}(2007)}]{Cornish:2006ms}
\bibinfo{author}{\bibfnamefont{N.~J.} \bibnamefont{Cornish}} \bibnamefont{and}
  \bibinfo{author}{\bibfnamefont{E.~K.} \bibnamefont{Porter}},
  \bibinfo{journal}{Class. Quant. Grav.} \textbf{\bibinfo{volume}{24}},
  \bibinfo{pages}{5729} (\bibinfo{year}{2007}), \eprint{gr-qc/0612091}.

\bibitem[{\citenamefont{Sokal}(1997)}]{dewitt-morette_monte_1997}
\bibinfo{author}{\bibfnamefont{A.}~\bibnamefont{Sokal}}, in
  \emph{\bibinfo{booktitle}{Functional {{Integration}}}}, edited by
  \bibinfo{editor}{\bibfnamefont{C.}~\bibnamefont{{DeWitt-Morette}}},
  \bibinfo{editor}{\bibfnamefont{P.}~\bibnamefont{Cartier}}, \bibnamefont{and}
  \bibinfo{editor}{\bibfnamefont{A.}~\bibnamefont{Folacci}}
  (\bibinfo{publisher}{{Springer US}}, \bibinfo{address}{{Boston, MA}},
  \bibinfo{year}{1997}), vol. \bibinfo{volume}{361}, pp.
  \bibinfo{pages}{131--192}, ISBN \bibinfo{isbn}{978-1-4899-0321-1
  978-1-4899-0319-8}.

\bibitem[{\citenamefont{Madras and Sokal}(1988)}]{Madras:1988ei}
\bibinfo{author}{\bibfnamefont{N.}~\bibnamefont{Madras}} \bibnamefont{and}
  \bibinfo{author}{\bibfnamefont{A.~D.} \bibnamefont{Sokal}},
  \bibinfo{journal}{J. Statist. Phys.} \textbf{\bibinfo{volume}{50}},
  \bibinfo{pages}{109} (\bibinfo{year}{1988}).

\bibitem[{\citenamefont{Dietrich et~al.}(2017)\citenamefont{Dietrich, Bernuzzi,
  and Tichy}}]{Dietrich:2017aum}
\bibinfo{author}{\bibfnamefont{T.}~\bibnamefont{Dietrich}},
  \bibinfo{author}{\bibfnamefont{S.}~\bibnamefont{Bernuzzi}}, \bibnamefont{and}
  \bibinfo{author}{\bibfnamefont{W.}~\bibnamefont{Tichy}},
  \bibinfo{journal}{Phys. Rev. D} \textbf{\bibinfo{volume}{96}},
  \bibinfo{pages}{121501} (\bibinfo{year}{2017}), \eprint{1706.02969}.

\bibitem[{\citenamefont{Dietrich et~al.}(2019)\citenamefont{Dietrich, Samajdar,
  Khan, Johnson-McDaniel, Dudi, and Tichy}}]{Dietrich:2019kaq}
\bibinfo{author}{\bibfnamefont{T.}~\bibnamefont{Dietrich}},
  \bibinfo{author}{\bibfnamefont{A.}~\bibnamefont{Samajdar}},
  \bibinfo{author}{\bibfnamefont{S.}~\bibnamefont{Khan}},
  \bibinfo{author}{\bibfnamefont{N.~K.} \bibnamefont{Johnson-McDaniel}},
  \bibinfo{author}{\bibfnamefont{R.}~\bibnamefont{Dudi}}, \bibnamefont{and}
  \bibinfo{author}{\bibfnamefont{W.}~\bibnamefont{Tichy}},
  \bibinfo{journal}{Phys. Rev. D} \textbf{\bibinfo{volume}{100}},
  \bibinfo{pages}{044003} (\bibinfo{year}{2019}), \eprint{1905.06011}.

\bibitem[{\citenamefont{Husa et~al.}(2016)\citenamefont{Husa, Khan, Hannam,
  P\"urrer, Ohme, Forteza, and Boh\'e}}]{PhysRevD.93.044006}
\bibinfo{author}{\bibfnamefont{S.}~\bibnamefont{Husa}},
  \bibinfo{author}{\bibfnamefont{S.}~\bibnamefont{Khan}},
  \bibinfo{author}{\bibfnamefont{M.}~\bibnamefont{Hannam}},
  \bibinfo{author}{\bibfnamefont{M.}~\bibnamefont{P\"urrer}},
  \bibinfo{author}{\bibfnamefont{F.}~\bibnamefont{Ohme}},
  \bibinfo{author}{\bibfnamefont{X.~J.} \bibnamefont{Forteza}},
  \bibnamefont{and} \bibinfo{author}{\bibfnamefont{A.}~\bibnamefont{Boh\'e}},
  \bibinfo{journal}{Phys. Rev. D} \textbf{\bibinfo{volume}{93}},
  \bibinfo{pages}{044006} (\bibinfo{year}{2016}).

\bibitem[{\citenamefont{Khan et~al.}(2016)\citenamefont{Khan, Husa, Hannam,
  Ohme, P\"urrer, Jim\'enez~Forteza, and Boh\'e}}]{Khan:2015jqa}
\bibinfo{author}{\bibfnamefont{S.}~\bibnamefont{Khan}},
  \bibinfo{author}{\bibfnamefont{S.}~\bibnamefont{Husa}},
  \bibinfo{author}{\bibfnamefont{M.}~\bibnamefont{Hannam}},
  \bibinfo{author}{\bibfnamefont{F.}~\bibnamefont{Ohme}},
  \bibinfo{author}{\bibfnamefont{M.}~\bibnamefont{P\"urrer}},
  \bibinfo{author}{\bibfnamefont{X.}~\bibnamefont{Jim\'enez~Forteza}},
  \bibnamefont{and} \bibinfo{author}{\bibfnamefont{A.}~\bibnamefont{Boh\'e}},
  \bibinfo{journal}{Phys. Rev. D} \textbf{\bibinfo{volume}{93}},
  \bibinfo{pages}{044007} (\bibinfo{year}{2016}), \eprint{1508.07253}.

\bibitem[{\citenamefont{Apostolatos et~al.}(1994)\citenamefont{Apostolatos,
  Cutler, Sussman, and Thorne}}]{Apostolatos:1994mx}
\bibinfo{author}{\bibfnamefont{T.~A.} \bibnamefont{Apostolatos}},
  \bibinfo{author}{\bibfnamefont{C.}~\bibnamefont{Cutler}},
  \bibinfo{author}{\bibfnamefont{G.~J.} \bibnamefont{Sussman}},
  \bibnamefont{and} \bibinfo{author}{\bibfnamefont{K.~S.}
  \bibnamefont{Thorne}}, \bibinfo{journal}{Phys. Rev. D}
  \textbf{\bibinfo{volume}{49}}, \bibinfo{pages}{6274} (\bibinfo{year}{1994}).

\bibitem[{\citenamefont{Ajith et~al.}(2011)}]{Ajith:2009bn}
\bibinfo{author}{\bibfnamefont{P.}~\bibnamefont{Ajith}} \bibnamefont{et~al.},
  \bibinfo{journal}{Phys. Rev. Lett.} \textbf{\bibinfo{volume}{106}},
  \bibinfo{pages}{241101} (\bibinfo{year}{2011}), \eprint{0909.2867}.

\bibitem[{\citenamefont{Santamaria et~al.}(2010)}]{Santamaria:2010yb}
\bibinfo{author}{\bibfnamefont{L.}~\bibnamefont{Santamaria}}
  \bibnamefont{et~al.}, \bibinfo{journal}{Phys. Rev. D}
  \textbf{\bibinfo{volume}{82}}, \bibinfo{pages}{064016}
  (\bibinfo{year}{2010}), \eprint{1005.3306}.

\bibitem[{\citenamefont{Damour}(2001)}]{Damour:2001tu}
\bibinfo{author}{\bibfnamefont{T.}~\bibnamefont{Damour}},
  \bibinfo{journal}{Phys. Rev. D} \textbf{\bibinfo{volume}{64}},
  \bibinfo{pages}{124013} (\bibinfo{year}{2001}), \eprint{gr-qc/0103018}.

\bibitem[{\citenamefont{Flanagan and Hinderer}(2008)}]{Flanagan:2007ix}
\bibinfo{author}{\bibfnamefont{E.~E.} \bibnamefont{Flanagan}} \bibnamefont{and}
  \bibinfo{author}{\bibfnamefont{T.}~\bibnamefont{Hinderer}},
  \bibinfo{journal}{Phys. Rev. D} \textbf{\bibinfo{volume}{77}},
  \bibinfo{pages}{021502} (\bibinfo{year}{2008}), \eprint{0709.1915}.

\bibitem[{\citenamefont{Wade et~al.}(2014)\citenamefont{Wade, Creighton,
  Ochsner, Lackey, Farr, Littenberg, and Raymond}}]{Wade:2014vqa}
\bibinfo{author}{\bibfnamefont{L.}~\bibnamefont{Wade}},
  \bibinfo{author}{\bibfnamefont{J.~D.~E.} \bibnamefont{Creighton}},
  \bibinfo{author}{\bibfnamefont{E.}~\bibnamefont{Ochsner}},
  \bibinfo{author}{\bibfnamefont{B.~D.} \bibnamefont{Lackey}},
  \bibinfo{author}{\bibfnamefont{B.~F.} \bibnamefont{Farr}},
  \bibinfo{author}{\bibfnamefont{T.~B.} \bibnamefont{Littenberg}},
  \bibnamefont{and} \bibinfo{author}{\bibfnamefont{V.}~\bibnamefont{Raymond}},
  \bibinfo{journal}{Phys. Rev. D} \textbf{\bibinfo{volume}{89}},
  \bibinfo{pages}{103012} (\bibinfo{year}{2014}), \eprint{1402.5156}.

\bibitem[{\citenamefont{Favata}(2014)}]{Favata:2013rwa}
\bibinfo{author}{\bibfnamefont{M.}~\bibnamefont{Favata}},
  \bibinfo{journal}{Phys. Rev. Lett.} \textbf{\bibinfo{volume}{112}},
  \bibinfo{pages}{101101} (\bibinfo{year}{2014}), \eprint{1310.8288}.

\bibitem[{\citenamefont{Lange et~al.}(2018{\natexlab{b}})\citenamefont{Lange,
  O'Shaughnessy, and Rizzo}}]{Lange:2018pyp}
\bibinfo{author}{\bibfnamefont{J.}~\bibnamefont{Lange}},
  \bibinfo{author}{\bibfnamefont{R.}~\bibnamefont{O'Shaughnessy}},
  \bibnamefont{and} \bibinfo{author}{\bibfnamefont{M.}~\bibnamefont{Rizzo}}
  (\bibinfo{year}{2018}{\natexlab{b}}), \eprint{1805.10457}.

\bibitem[{\citenamefont{Chollet et~al.}(2015)}]{chollet2015keras}
\bibinfo{author}{\bibfnamefont{F.}~\bibnamefont{Chollet}} \bibnamefont{et~al.},
  \emph{\bibinfo{title}{Keras}}, \bibinfo{howpublished}{\url{https://keras.io}}
  (\bibinfo{year}{2015}).

\bibitem[{\citenamefont{Abadi et~al.}(2015)\citenamefont{Abadi, Agarwal,
  Barham, Brevdo, Chen, Citro, Corrado, Davis, Dean, Devin
  et~al.}}]{tensorflow2015-whitepaper}
\bibinfo{author}{\bibfnamefont{M.}~\bibnamefont{Abadi}},
  \bibinfo{author}{\bibfnamefont{A.}~\bibnamefont{Agarwal}},
  \bibinfo{author}{\bibfnamefont{P.}~\bibnamefont{Barham}},
  \bibinfo{author}{\bibfnamefont{E.}~\bibnamefont{Brevdo}},
  \bibinfo{author}{\bibfnamefont{Z.}~\bibnamefont{Chen}},
  \bibinfo{author}{\bibfnamefont{C.}~\bibnamefont{Citro}},
  \bibinfo{author}{\bibfnamefont{G.~S.} \bibnamefont{Corrado}},
  \bibinfo{author}{\bibfnamefont{A.}~\bibnamefont{Davis}},
  \bibinfo{author}{\bibfnamefont{J.}~\bibnamefont{Dean}},
  \bibinfo{author}{\bibfnamefont{M.}~\bibnamefont{Devin}},
  \bibnamefont{et~al.}, \emph{\bibinfo{title}{{TensorFlow}: Large-scale machine
  learning on heterogeneous systems}} (\bibinfo{year}{2015}),
  \bibinfo{note}{software available from tensorflow.org},
  \urlprefix\url{https://www.tensorflow.org/}.

\bibitem[{\citenamefont{Srivastava et~al.}(2014)\citenamefont{Srivastava,
  Hinton, Krizhevsky, Sutskever, and Salakhutdinov}}]{srivastava_dropout_2014}
\bibinfo{author}{\bibfnamefont{N.}~\bibnamefont{Srivastava}},
  \bibinfo{author}{\bibfnamefont{G.}~\bibnamefont{Hinton}},
  \bibinfo{author}{\bibfnamefont{A.}~\bibnamefont{Krizhevsky}},
  \bibinfo{author}{\bibfnamefont{I.}~\bibnamefont{Sutskever}},
  \bibnamefont{and}
  \bibinfo{author}{\bibfnamefont{R.}~\bibnamefont{Salakhutdinov}},
  \bibinfo{journal}{Journal of Machine Learning Research}
  \textbf{\bibinfo{volume}{15}}, \bibinfo{pages}{1929} (\bibinfo{year}{2014}),
  ISSN \bibinfo{issn}{1533-7928},
  \urlprefix\url{http://jmlr.org/papers/v15/srivastava14a.html}.

\bibitem[{\citenamefont{Geron}(2019)}]{10.5555/3378999}
\bibinfo{author}{\bibfnamefont{A.}~\bibnamefont{Geron}},
  \emph{\bibinfo{title}{Hands-On Machine Learning with Scikit-Learn, Keras, and
  TensorFlow: Concepts, Tools, and Techniques to Build Intelligent Systems}}
  (\bibinfo{publisher}{O'Reilly Media, Inc.}, \bibinfo{year}{2019}),
  \bibinfo{edition}{2nd} ed., ISBN \bibinfo{isbn}{1492032646}.

\bibitem[{\citenamefont{Pedregosa et~al.}(2011)\citenamefont{Pedregosa,
  Varoquaux, Gramfort, Michel, Thirion, Grisel, Blondel, Prettenhofer, Weiss,
  Dubourg et~al.}}]{scikit-learn}
\bibinfo{author}{\bibfnamefont{F.}~\bibnamefont{Pedregosa}},
  \bibinfo{author}{\bibfnamefont{G.}~\bibnamefont{Varoquaux}},
  \bibinfo{author}{\bibfnamefont{A.}~\bibnamefont{Gramfort}},
  \bibinfo{author}{\bibfnamefont{V.}~\bibnamefont{Michel}},
  \bibinfo{author}{\bibfnamefont{B.}~\bibnamefont{Thirion}},
  \bibinfo{author}{\bibfnamefont{O.}~\bibnamefont{Grisel}},
  \bibinfo{author}{\bibfnamefont{M.}~\bibnamefont{Blondel}},
  \bibinfo{author}{\bibfnamefont{P.}~\bibnamefont{Prettenhofer}},
  \bibinfo{author}{\bibfnamefont{R.}~\bibnamefont{Weiss}},
  \bibinfo{author}{\bibfnamefont{V.}~\bibnamefont{Dubourg}},
  \bibnamefont{et~al.}, \bibinfo{journal}{Journal of Machine Learning Research}
  \textbf{\bibinfo{volume}{12}}, \bibinfo{pages}{2825} (\bibinfo{year}{2011}).

\bibitem[{\citenamefont{{Kingma} and {Ba}}(2014)}]{2014arXiv1412.6980K}
\bibinfo{author}{\bibfnamefont{D.~P.} \bibnamefont{{Kingma}}} \bibnamefont{and}
  \bibinfo{author}{\bibfnamefont{J.}~\bibnamefont{{Ba}}},
  \bibinfo{journal}{arXiv e-prints} \bibinfo{eid}{arXiv:1412.6980}
  (\bibinfo{year}{2014}), \eprint{1412.6980}.

\bibitem[{\citenamefont{{Masters} and {Luschi}}(2018)}]{2018arXiv180407612M}
\bibinfo{author}{\bibfnamefont{D.}~\bibnamefont{{Masters}}} \bibnamefont{and}
  \bibinfo{author}{\bibfnamefont{C.}~\bibnamefont{{Luschi}}},
  \bibinfo{journal}{arXiv e-prints} \bibinfo{eid}{arXiv:1804.07612}
  (\bibinfo{year}{2018}), \eprint{1804.07612}.

\end{thebibliography}
\onecolumngrid
\appendix

\section{Deriving the log-prior gradients with respect to $(\lnmc, \lnmu)$}\label{sec:app_mass_prior}
The LVK analysis assumes a flat prior in the individual masses, i.e. $\pi(m_1, m_2) = \pi(m_1)\pi(m_2)\propto 1$.  To transform for the joint prior in 
individual masses to the joint prior in chirp and reduced masses, we begin with the expression
\beq
\pi(\ln\mc, \ln\m) = J\, \pi(m_1, m_2),
\eeq
where $J$ is the Jacobian determinant
\beq
J = \left| \frac{\partial(m_1, m_2)}{\partial(\ln\mc, \ln\m)}\right|.
\eeq
Using the chain rule, we can write
\beq
J = J_1 J_2 =  \left| \frac{\partial(m_1, m_2)}{\partial(\mc, \m)}\right|  \left| \frac{\partial(\mc, \m)}{\partial(\ln\mc, \ln\m)}\right|. 
\eeq
For ease of calculation, it is easier to work with the inverse of the Jacobian determinants, i.e. $J_i^{-1}$.  Expanding $J_1^{-1}$, we get
\beq
J_1^{-1} = \left| \frac{\partial(\mc, \m)}{\partial(m_1, m_2)}\right| = \left(\frac{\partial \mc}{\partial m_1}\frac{\partial\m}{\partial m_2}-\frac{\partial\mc}{\partial m_2}\frac{\partial \m}{\partial m_1}\right)^{-1}.
\eeq
Here we find
\bea
\frac{\partial \mc}{\partial m_1} &=& \frac{m_2^{3/5} (2m_1 + 3m_2)}{5 m_1^{2/5}(m_1+m_2)^{6/5}},\\
\frac{\partial \m}{\partial m_1} &=& \frac{m_2^2}{(m_1+m_2)^2},\\
\frac{\partial \mc}{\partial m_2} &=& \frac{m_1^{3/5} (3m_1 + 2m_2)}{5 m_2^{2/5}(m_1+m_2)^{6/5}},\\
\frac{\partial \m}{\partial m_2} &=& \frac{m_1^2}{(m_1+m_2)^2}.\\
\eea
Combining these expressions gives
\beq
J_1^{-1} = \frac{m_1^{8/5}m_2^{3/5} (2m_1 + 3m_2)}{5 (m_1+m_2)^{16/5}} - \frac{m_1^{3/5} m_2^{8/5}(3m_1 + 2m_2)}{5 (m_1+m_2)^{16/5}}.
\eeq
Simplifying the above expression and re-inverting, we obtain
\beq
J_1  =  \left| \frac{\partial(m_1, m_2)}{\partial(\mc, \m)}\right| = \frac{5(m_1+m_2)^{11/5}}{2 (m_1 m_2)^{3/5}(m_1 - m_2)}.
\eeq
Similarly, expanding $J_2^{-1}$, we get
\beq
J_2^{-1} = \left| \frac{\partial(\mc, \m)}{\partial(\ln\mc, \ln\m)}\right| = \left(\frac{\partial\ln \mc}{\partial \mc}\frac{\partial\ln\m}{\partial \m}\right)^{-1},
\eeq
where
\bea
\frac{\partial\ln \mc}{\partial \mc} &=& \mc^{-1},\\
\frac{\partial\ln\m}{\partial \m} &=& \m^{-1},
\eea
giving
\beq
J_2 = \left| \frac{\partial(\mc, \m)}{\partial(\ln\mc, \ln\m)}\right| = \mc\m.
\eeq
This allows us to finally express the full Jacobian determinant as
\beq
J = \left| \frac{\partial(m_1, m_2)}{\partial(\ln\mc, \ln\m)}\right| = \frac{5(m_1+m_2)^{11/5}\mc\m}{2 (m_1 m_2)^{3/5}(m_1 - m_2)}.
\eeq

The next step is to write this expression completely in terms of $(\mc,\m)$.  To begin with, we use the identities $(m_1+m_2) = \mc^{5/2}/\m^{3/2}$
and $\eta = (\m/\mc)^{5/2}$.  This allows us to write
\beq
m_1 - m_2 = m\sqrt{1-4\eta} = \frac{\mc^{5/2}}{\m^{3/2}}\sqrt{1-4\left(\frac{\m}{\mc}\right)^{5/2}}.
\eeq
Furthermore, we can write
\beq
m_1 m_2 = m^2\eta = \frac{\mc^5 \m^{5/2}}{\m^3 \mc^{5/2}} = \frac{\mc^{5/2}}{\sqrt{\m}}.
\eeq
Defining a function
\beq
x = 1-4\left(\frac{\m}{\mc}\right)^{5/2},
\eeq
and substituting the above terms in the expression for the Jacobian, we arrive at the expression
\beq
J(\mc, \m) = \frac{5\mc^{5/2}}{2\sqrt{\m}\sqrt{ x}},
\eeq
allowing us to write
\beq
\pi(\ln\mc, \ln\m) =  \frac{5\mc^{5/2}}{2\sqrt{\m}\sqrt{ x}}\, \pi(m_1, m_2).
\eeq

To calculate the gradient of the log-prior wrt $\ln\mc$, we write
\beq
\frac{\partial\ln\pi(\ln\mc, \ln\m)}{\partial\ln\mc} = \frac{\partial \ln J}{\partial\ln\mc} + \frac{\partial\ln\pi(m_1)}{\partial\ln\mc} + \frac{\partial\ln\pi(m_2)}{\partial\ln\mc}.
\eeq
As the prior distributions for the individual masses are uniform, the second and third terms are zero.  To simplify calculations, we proceed by writing
\beq
\frac{\partial \ln J}{\partial\ln\mc}=\frac{\mc}{J}\frac{\partial  J}{\partial\mc}.
\eeq
With some algebra, we can derive
\beq
\frac{\partial  J}{\partial\mc} = \frac{25(\mc^{5/2}x - 2\m^{5/2})}{4\sqrt{\m}\mc x^{3/2}},
\eeq
giving
\beq
\frac{\partial\ln\pi(\ln\mc, \ln\m)}{\partial\ln\mc} = \frac{5(\mc^{5/2}x - 2\m^{5/2})}{2\mc^{5/2}x}.
\eeq
Similarly, by writing
\beq
\frac{\partial \ln J}{\partial\ln\m}=\frac{\m}{J}\frac{\partial  J}{\partial\m},
\eeq
we can express
\beq
\frac{\partial  J}{\partial\m} = -\frac{5(\mc^{5/2}x^{3/2} - 10\m^{2}x^{1/2})}{4\m^{3/2} x^{2}},
\eeq
allowing us to write
\beq
\frac{\partial\ln\pi(\ln\mc, \ln\m)}{\partial\ln\m}= -\frac{(\mc^{5/2}x - 10\m^{2})}{2\mc^{5/2}x}.
\eeq

\section{$R^2$ scores against a test data set}\label{sec:R2_scores}
Here we present typical $R^2$ scores after training the DNN model.  At the end of Phase I of the algorithm, a data set of 666,000 data points is produced.  This data set is divided into three parts: training ($70\%$), validation ($20\%$) and test ($10\%$) 
data sets.  To evaluate the DNN model, we compare the DNN-model gradients calculated at the 66,000 test data points against the numerical gradients from the same point.  We set our performance threshold at $R^2 > 0.99$ (implying that 99\% of the data is fit by the model) or 500 training epochs for unimodal
sources, and $R^2 > 0.9$ (implying that 90\% of the data is fit by the model) or 1000 epochs for multimodal sources.

\begin{figure*}[th]
       \includegraphics[width=\textwidth, height=9cm]{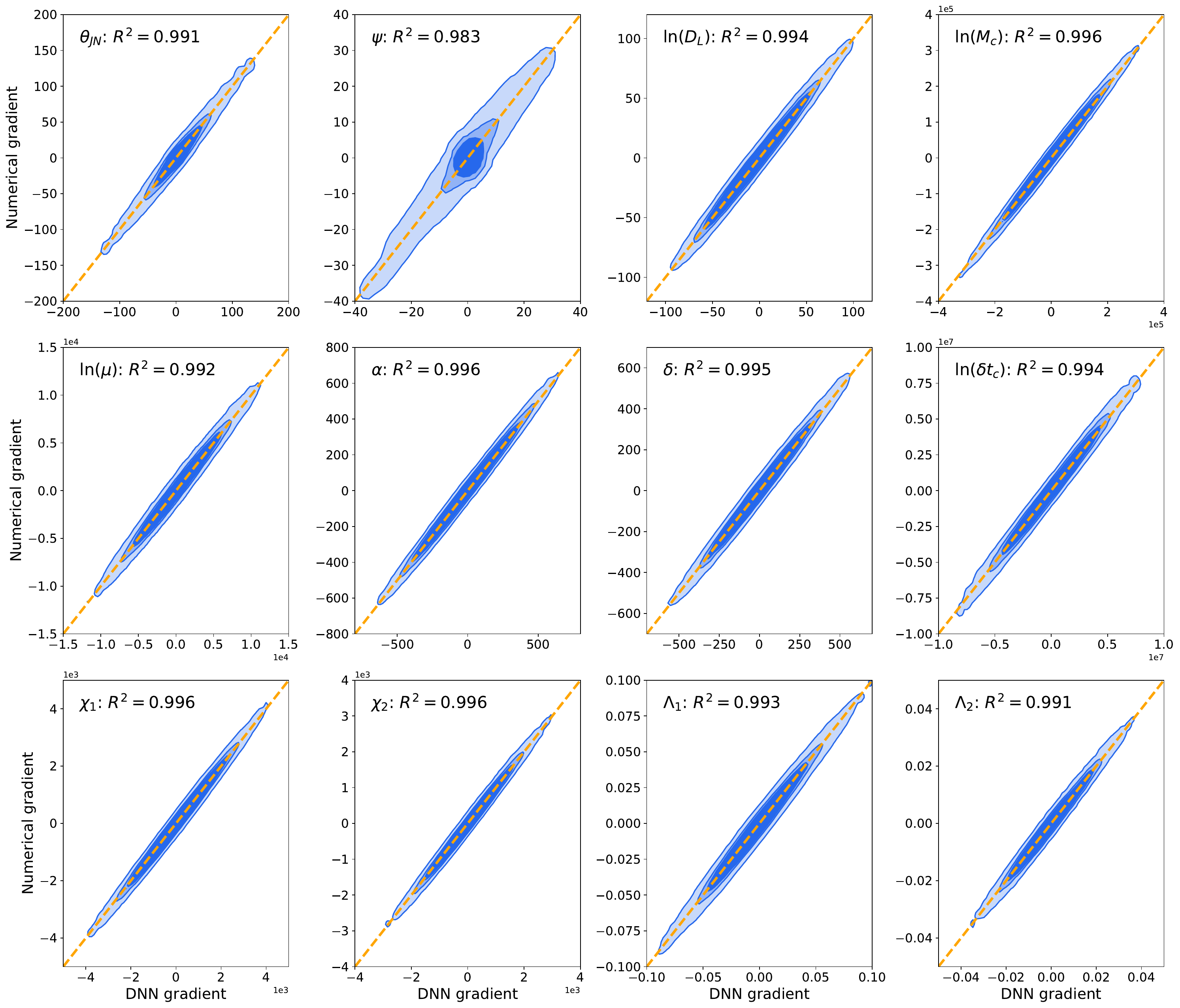}
        \caption{A comparison between gradients of the log-likelihood using numerical differencing and the DNN approximation on a test set of 66,000 data points in the 12D parameter space for GW170817.  The threshold for unimodal sources is set at $R^2 > 0.99$ or 500 training epochs..}
        \label{fig:GW170817_R2}.   
\end{figure*}

\begin{figure*}[th]
       \includegraphics[width=\textwidth, height=9cm]{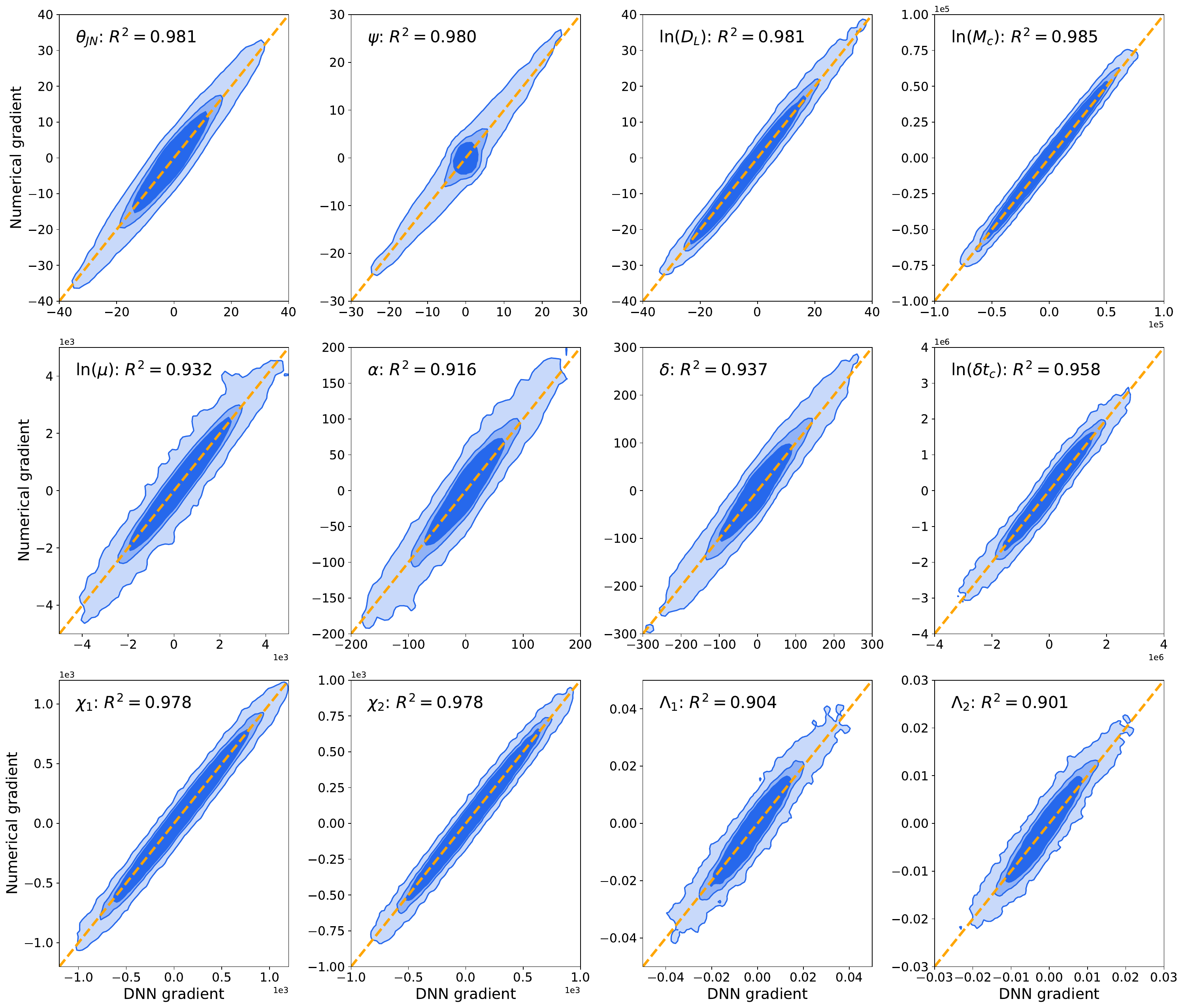}
        \caption{A comparison between gradients of the log-likelihood using numerical differencing and the DNN approximation on a test set of 66,000 data points in the 12D parameter space for GW190425.  The threshold for multimodal sources is set at $R^2 > 0.9$ or 1000 training epochs.}
        \label{fig:GW190425_R2}.   
\end{figure*}

\end{document}